\documentclass{article}

\usepackage{arxiv}

\usepackage[utf8]{inputenc} 
\usepackage[T1]{fontenc}    
\usepackage{hyperref}       
\usepackage{url}            
\usepackage{booktabs}       
\usepackage{amsfonts}       
\usepackage{nicefrac}       
\usepackage{microtype}      
\usepackage{natbib}
\usepackage{doi}

\usepackage{revsymb}
\usepackage{float}
\usepackage{mathtools,amssymb,xfrac}
\usepackage{amsmath}
\usepackage{mathrsfs}
\usepackage{color}
\usepackage{rotating}
\usepackage{graphicx,epsfig,subfigure}
\usepackage{array}
\usepackage{tabularx,multirow}

\title{Systematic analysis reveals key microRNAs as diagnostic and prognostic factors in progressive stages of lung cancer}

\author{Dietrich Kong \dag\\
Department of Computational Physics \quad \\
	Institute of Modern Physics\\
	Chinese Academy of Sciences \\
	\texttt{dietrichkong@gmail.com} \\
	\And
	Ke Wang \dag\\
	Department of Computational Physics \\
	Institute of Modern Physics\\
	Chinese Academy of Sciences\\
	\texttt{kvickywang@gmail.com} \\
	\And
	Qiu-Ning Zhang \\
	\quad Radiation medicine laboratory \quad \\
	Institute of Modern Physics\\
	Chinese Academy of Sciences \\
	\texttt{zhangqn@impcas.ac.cn}\\
	\And
	\href{https://orcid.org/0000-0002-6302-8612}{\includegraphics[scale=0.06]{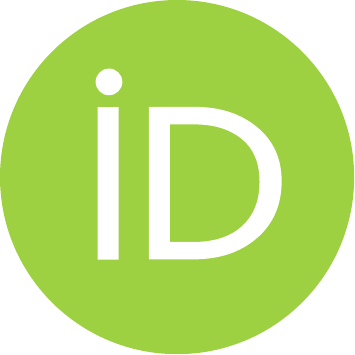}\hspace{1mm}Zhi-Tong Bing}\thanks{Corresponding Author. {Email: bingzt@impcas.ac.cn}}\\
	Department of Computational Physics\\
	Institute of Modern Physics\\
	Chinese Academy of Sciences\\
	Advanced Energy Science and Technology\\
	Guangdong Laboratory\\
	\texttt{bingzt@impcas.ac.cn} \\
}

\date{}

\renewcommand{\headeright}{Dietrich et al}
\renewcommand{\undertitle}{~}
\renewcommand{\shorttitle}{Diagnostic and prognostic microRNAs in lung cancer}

\hypersetup{
pdftitle={Diagnostic and prognostic microRNAs in lung cancer},
pdfsubject={Computational Biology},
pdfauthor={Dietrich et al.},
pdfkeywords={MicroRNA, Machine learning, Clinical information, Lung cancer},
}

\begin{document}
\maketitle

\begin{abstract}
	MicroRNAs play an indispensable role in numerous biological processes 
ranging from organismic development to tumor progression. 
In oncology, these microRNAs constitute a fundamental regulation 
role in the pathology of cancer that provides the basis for 
probing into the influences on clinical features 
through transcriptome data. 
Previous work focused on machine learning (ML) for 
searching biomarkers in different cancer databases, but 
the functions of these biomarkers are fully not clear. 
Taking lung cancer as a prototype case of study.
Through integrating clinical information into the transcripts 
expression data, we systematically analyzed the effect of microRNA on 
diagnostic and prognostic factors at deteriorative 
lung adenocarcinoma (LUAD). 
After dimension reduction, unsupervised hierarchical clustering 
was used to find the diagnostic factors which represent the unique 
expression patterns of microRNA at various patient's stages.
In addition, we developed a classification framework, 
Light Gradient Boosting Machine (LightGBM) and 
SHAPley Additive explanation (SHAP) algorithm, to screen out 
the prognostic factors. 
Enrichment analyses show that the diagnostic and prognostic factors 
are not only enriched in cancer-related pathways, but also involved 
in many vital cellular signaling transduction and immune responses.
These key microRNAs also impact the survival risk of LUAD patients 
at all (or a specific) stage(s) and some of them target 
some important Transcription Factors (TF). 
The key finding is that five microRNAs (hsa-mir-196b, hsa-mir-31, 
hsa-mir-891a, hsa-mir-34c, and hsa-mir-653) can then serve as not 
only potential diagnostic factors but also prognostic tools in 
the monitoring of lung cancer.
\end{abstract}

\keywords{MicroRNA \and Lung cancer \and Machine learning \and Clinical information}

\section{Introduction}
Lung cancer is the leading cause of cancer-related human deaths
worldwide~\citep{SFSLSJB:2021}. Approximately 85$\%$ of the lung cancer cases
can be classified as non small-cell lung cancer (NSCLC), among which lung
adenocarcinoma (LUAD) is one of the most common subtypes~\citep{SMJ:2018}.
The molecular mechanisms behind cancer evolution are extremely complex,
impeding accurate and reliable and diagnosis and prognosis as well as effective
treatment~\citep{MYCSA:2008}. A milestone discovery the general developmental 
and disease contexts is the roles played by non-coding RNAs (ncRNAs)~\citep{AJS:2018}, 
especially microRNAs~\citep{SGBMAKK:2021,B2018,GM2019,CHWYS2019}. 
In in cancer research, at the RNA level, the microRNA has been found in key 
regulators of physiological functions and major cancer types
~\citep{ABSSR2020,WGLCW2018,LLZ2019,HCHWC2020,ZMZCL2021}.
Particularly relevant to lung cancer, microRNA had been identified 
as the oncogenic drivers and tumor suppressors
~\citep{WTLKH:2019,YCGHZ2018,NYGSM:2020,PMSK:2021}. 

A standard existing approach to monitoring tumor progress and 
detecting/ascertaining the underlying mechanism is non-coding 
RNA transcriptomics, which has led to a large number of significant
biomarkers and therapeutic targets~\citep{Seoetal:2012,CGARN:2014,EPSI:2019,DCFCK2021}. 
In previous studies of ncRNA, a commonly practiced methodology
is to identify some Differentially Expressed (DE) RNAs
based on absolute Fold Changes (FC) and values of the false positive ratio
~\citep{YYKCJ2019,ZGZB:2020,YYSQ:2020}.
Some recent works input ncRNAs expression data into machine 
learning (ML) algorithms~\citep{LMMST2019,F2021,MGMYS2020}, such as LightGBM 
to seek markers. And others analyze how microRNA affects the prognosis 
of patients by combining clinical data~\citep{AKMYY2020,ZGZO2021,XCZLQ2020}.

These methods and analyses have inferred some microRNAs related 
to lung cancer and have been well verified, but there 
has been little systematic discussion of the relationship
between diagnostic factors and prognostic factors.
Nonetheless, it is unlikely that the intrinsic molecular remain 
static during cancer development, the dynamical interplay among 
various prognosis and diagnosis functions of RNAs was completely 
ignored. To better understand cancer and to identify more effective 
factors, the dynamical aspects of the ncRNA in LUAD must be 
taken into account. To remedy this deficiency had motivated our work. 
With the presently available gene expression and clinical information, we were able 
to incorporate investigation of diagnosis and prognosis of lung cancer.
In particular, we focused on the three stages of LUAD progression 
based on both transcriptome and corresponding clinical 
data of LUAD from The Cancer Genome Atlas~\footnote{https://portal.gdc.cancer.gov/}. 

Through selecting the unique pattern from hierarchical clustering 
of microRNA expression in each given stage of LUAD, we found that 
some microRNAs were potentially served as diagnostic factors 
with cancer stages. And proved that the expression pattern of 
these diagnostic factors does not directly correspond to the 
molecular characteristics, such as gene expression, FC value, 
and even survival risk variables.
In addition, we classified the LUAD samples of each stage into 
two categories based on survival differences and employed an 
efficient and stable LightGBM algorithm to train the microRNA 
expression data with binary labels in the three stages 
of LUAD respectively. Then a SHAP algorithm was applied to 
explain the mechanism of the ML classifiers and verify key 
prognostic factors. 

Performing the Kyoto Encyclopedia of Genes and Genomes (KEGG) 
pathway enrichment analyses of those diagnosis or prognosis 
factors, we found that these factors are involved in a 
wide range of cancer-related pathways and related to important 
signaling translation pathways and immune processes.
Finally, we carried out K-M survival curve analysis and 
constructed the regulation network of factors and their 
targeted Transtript Factors (TF). 
Our work established a more comprehensive diagnosis-prognosis 
framework than previous ones, not only providing tools 
to probe more deeply into the mechanisms of cancer evolution 
than previously possible but also having the potential to
lead to more effective drug targets for LUAD as well as 
other types of cancer.

\section{Results}\label{sec2}
\subsection{Differential Expression (DE) Analysis and 
Confirmation of Biological Function}
\subsubsection{DE microRNAs}
As described in \textbf{Materials and Methods}, to explore 
there lationship between diagnosis and prognosis in evolutionary 
carcinoma, we first obtained the DE RNAs by comparing the 
microRNAs expression values in the LUAD samples with those 
from the normal samples. The amounts of samples and DE 
microRNAs in each stage were shown in Supplementary Table 1.

In the volcano map, Figure~\ref{fig:DE}A, 
some of the microRNAs with significant FC and P-value 
($-log10(P)>20$, $log_{2}|FC|>2$) have been 
widely reported, such as microRNAs hsa-mir-21, 
hsa-mir-9, etc.~\citep{Liuetal:2010a,Guindeetal:2018}.
However, DE analysis cannot select those molecules with smaller 
FC values, but they may play a vital role. 
We hoped to understand the molecular mechanism of evolutionary 
cancer not only from statistical indicators (FC value) of 
transcripts expression but also from other 
dimensions through systematical analysis. 
Therefore, a loose threshold was chosen in the process of DE 
analysis ($P-value<0.05$, $log_{2}|FC|>=1$) so as to 
avoid significant molecules being abandoned, and finally 
127, 130, and 131 microRNAs were selected in the early, middle, 
and later stages correspondingly, as shown in Figure~\ref{fig:DE}B.

\begin{figure*}[!ht]
\centering
\includegraphics[width=0.9\linewidth]{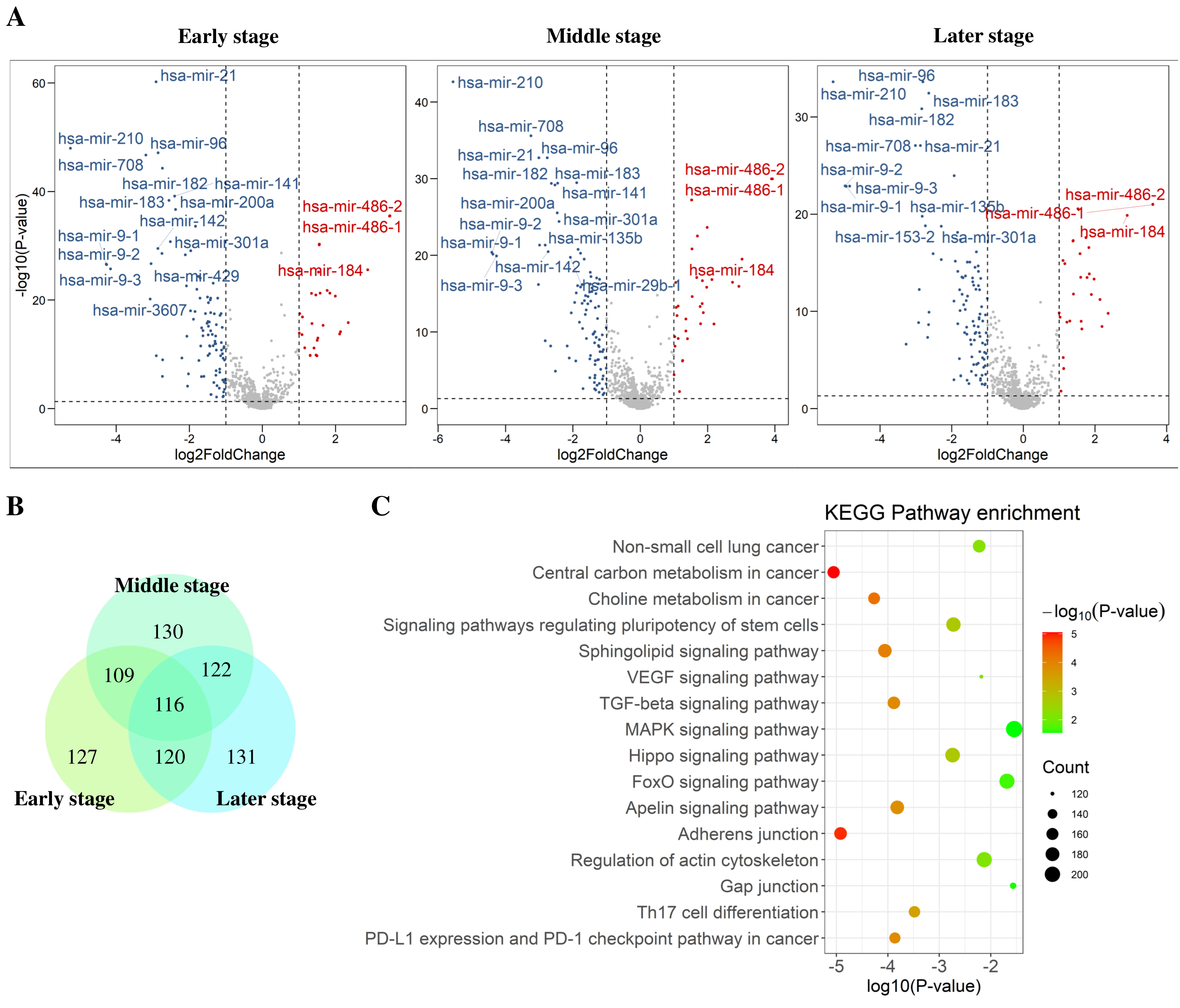}
\caption{\textbf{Differentially expressed microRNAs and their functions in LUAD}.
A. Volcano maps of the microRNA expression in early, middle, and later stages
of LUAD samples. The $x$-axis is $\log_{2}{\mbox{FC}}$ (FC value),
and the $y$-axis is the $-\log_{10}{\mbox{P}}$ from the DE analysis.
We selected the threshold $P<0.05$ and $\log_{2}{\mbox{FC}}>1$, especially, these microRNAs
which meet the threshold $-\log_{10}{\mbox{P}}>20$ and $\log_{2}{\mbox{FC}}>2$ 
were labeled by names.
B. The Venn diagram illustrates the number of DE microRNAs in different stages of LUAD,
reflecting molecules in the dynamical LUAD have a very high level of overlap.
C. The KEGG pathway enrichment analysis was performed on DE microRNAs. It revealed
that these microRNAs are closely related to the biological functions of the cancer
pathway, signal Translation, and immune pathway.} 
\label{fig:DE}
\end{figure*}

\subsubsection{Confirmation of Biological Function}
The current understanding of the biological functions of microRNAs 
is far less than coding protein messenger RNA, and little 
is known about the precursor microRNA. Considering that the content of 
precursor microRNA in organisms can be reflected in the amount of mature 
microRNAs in cells or tissues to a certain extent~\citep{LBA2016}. 
So the biological pathway enrichment analysis of mature microRNAs was used
to confirm the potential biological function of their precursor. 

According to the results of KEGG enrichment analysis in Figure~\ref{fig:DE}C, 
the DE precursor microRNAs are not only enriched 
in cancer-related pathways, such as ``MicroRNAs in cancer'', 
``Non-small cell lung cancer'' and ``Transcriptional misregulation 
in cancer''. It also involves many biological signaling pathways, 
such as Proteoglycans, Central carbon metabolism, Choline metabolism 
in cancer and the TGF-b/MAPK pathways. 
In addition, it plays a role in immune-related and other biological 
processes such as Th17 cell differentiation, cell adhesion, and motor protein. 
In addition, the microRNA enrichment pathways in different 
the various LUAD stages were also analyzed (shown in Supplementary Figure 1).

\subsection{Diagnostic Factors in Different LUAD Stages}
\subsubsection{The expression information being stored in the eigenvector}
In order to explore which of the DE microRNAs are suitable as 
diagnostic factors of the dynamical LUAD stages based on TNM classification
(see \textbf{Materials and Methods} for the meaning of TNM). 
We first reduced the dimensionality of the microRNA expression 
data to eliminate possible gene expression noise in the transcriptome 
data. The size of sample-microRNA expression matrices in the three 
stages of LUAD are (273, 127), (120, 130), and (103, 131), respectively. 
For these three matrices, we first decentralized it and calculated 
the covariance matrix, and then performed eigenvalue decomposition 
(\textbf{Materials and Methods} for more details). 
The 90\%$>$ information of microRNA expression data was retained, 
and finally, 30, 26, and 25 eigenvalues were selected in the early, 
middle, and later stages of LUAD patients.

\subsubsection{Eigenvector independent with expression and clinical characteristics}
The Pearson correlation coefficients between eigenvector 
weight and RNA expression or FC value are within plus 
or minus 0.3, as shown in Supplementary Figure 2.
Therefore, the numerical value of the eigenvector does not 
depend on the statistical indicators related to gene expression.
This analysis may ensure that the subsequent process to identify
the clinical function of microRNAs is all to a great extent independent 
of microRNA expression.

In addition, considering that the eigenvectors may be related 
to survival, rather than specific diagnostic factors of each stage, 
we employed the Cox proportional hazard regression model
(in Supplementary Note 2) to calculate the Hazard Risk of
eigenvectors and compare the two typical clinical 
characteristics, age and gender separately (in Supplementary Table 3). 
It's not the top eigenvectors that have the greatest
impact on survival ($p<0.05$ and $|HR-1|>0.1$), 
indicating that the data reduction process for finding diagnostic 
factors does not depend on patient survival information 
(in Supplementary Table 3). And thus the expression of DE 
microRNAs and survival data are statistically independent. 

\subsubsection{Unique expression patterns of microRNA}
Molecules served as a kind of diagnostic factors 
possess unique gene expression characteristics in each 
the gene expression of microRNAs should be potentially 
stages of LUAD, so the microRNAs with the distinguished pattern 
of the gene expression should be potentially 
considered as the stage-based diagnostic factors.
What's more, clustering by the value of the 
microRNA expression corresponding to the eigenvector 
load is beneficial for finding molecules with 
abnormal expression patterns in the data structure 
itself, rather than just directly looking for the gene 
expression characteristics.

\subsubsection{Hierarchical clustering of microRNAs expression}
\begin{figure*}[!ht]
\centering
\includegraphics[width=0.9\linewidth]{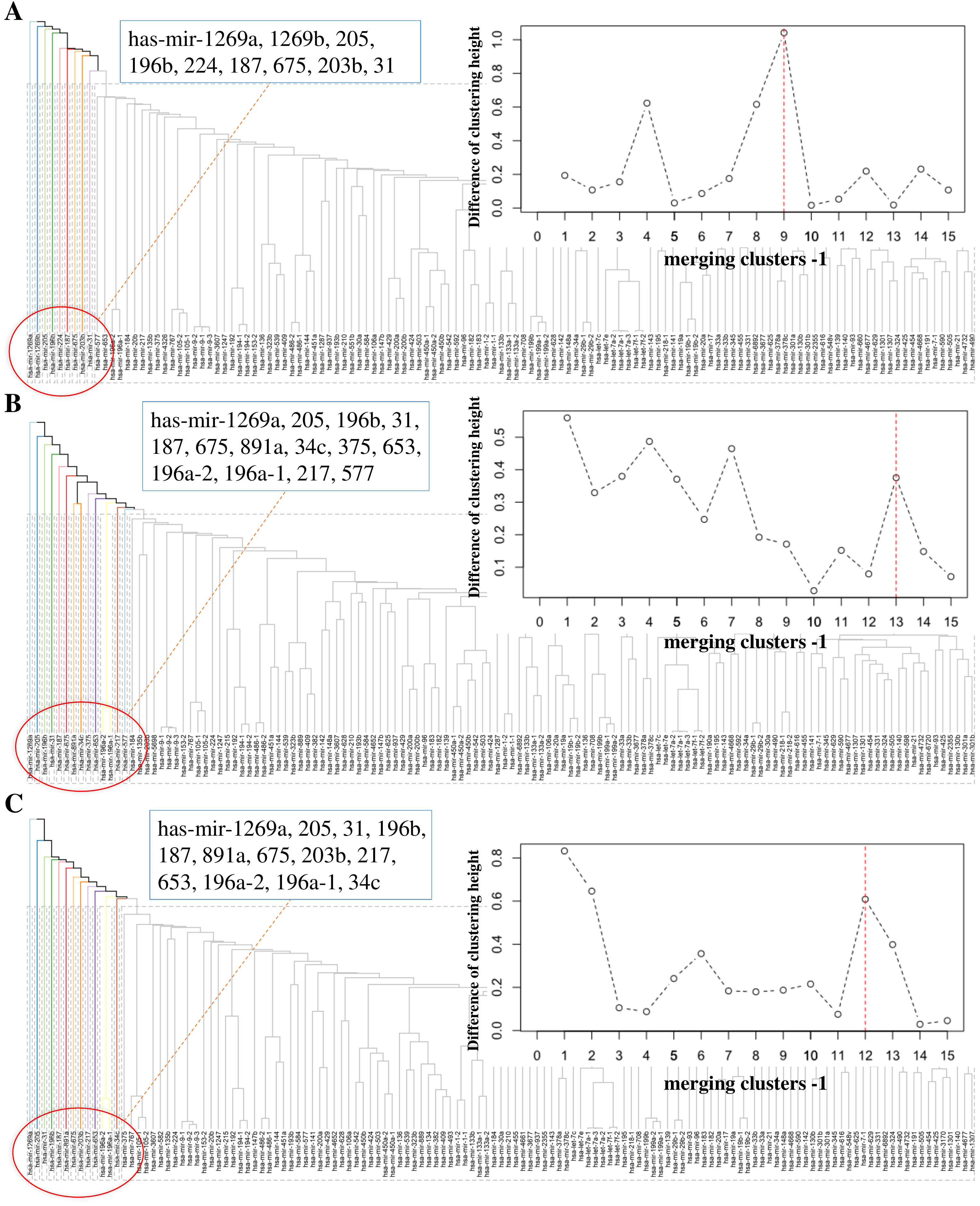}
\caption{\textbf{Clustering analysis for diagnostic factors of three LUAD stages}.
The unsupervised hierarchical clustering of microRNA expression eigenvectors 
was shown in A. early, B. middle, and C. later LUAD stages. 
Compared with the right part (gray) which is a group of microRNAs with 
a very similar expression, the color part on the left has a relatively 
more varied expression pattern, so these molecules have a
greater degree of changes of expression with each other.
The greater the distance from left to right in the clustering graph, 
the more different the expression patterns of ``sub-categories''. 
The line graph at the top-right depicted clustering heigh (spatial Euclidean distance) 
between each different microRNA cluster in a given 
stage of LUAD. The red dotted lines indicate that 
9, 13, and 12 are suitable criteria for the top 15 clusters.
The total number of microRNAs are 127, 130, and 134 in three stages respectively.
}
\label{fig:cluster}
\end{figure*}

Figure~\ref{fig:cluster} depicts the results of unsupervised 
hierarchical clustering of microRNAs expression
eigenvectors with three LUAD stages. Each molecule was 
classified into a group according to its Euclidean distance, 
and this process was repeated until all molecules 
are gathered into the same super-category. 
The molecular expression patterns within clusters are similar, 
while the differences between various clusters are large.
The greater the distance from left to right in the clustering 
diagram Figure~\ref{fig:cluster}, the more different the expression 
patterns of the sub-categories. The left sub-categories (colored) 
have a larger difference compared with the overall categories 
on the right (gray). 
Therefore, the gray part on the right side of the figure is 
a group of microRNAs with very similar 
expression patterns, while the left side is a number of 
molecules with relatively more varied expression patterns 
and thus those molecule(s) can be seen as a single category. 

In the meantime, we calculated clustering height 
which is a measure of discrimination for microRNA expression 
patterns, illustrated in the top-right line graphs 
(Figure~\ref{fig:cluster}). 
In particular, we ranked the spatial Euclidean distances between 
a cluster of microRNA(s) and other all clusters, and the largest 
difference of it was marked by the red dotted line which was used as the 
threshold. Finally, the top 9, 13, and 12 clusters were screened 
out in the early, middle, and later LUAD stages. 
Therefore, we extracted those microRNA with unique expression 
patterns in each LUAD stage separately (Figure~\ref{fig:cluster}). 

\subsubsection{Diagnostic factors with independent characteristics}
Finally, We screened out 9, 14, and 13 microRNAs (More details in The 
blue rectangular box in Figure~\ref{fig:cluster}) respectively 
that could be considered as diagnostic factors in early, middle, 
and later LUAD stages.
It is worth mentioning that these microRNAs are totally not similar 
with significant microRNA (~$-log10(P-value)>20$, ~$log_{2}|FC|>2$) 
from DE analysis at each LUAD stage. 
For example, the gene expression pattern of has-mir-1269a is distinctive 
from other molecules, but its expression and FC value 
(Figure~\ref{fig:DE}A) are not significant.

In short, by the dimensionality reduction and unsupervised 
hierarchical clustering, the microRNAs that we obtained with specific 
unique expression patterns in each stage, therefore, can be reasonably 
regarded as diagnostic factors which are independent from gene 
expression, FC Value, and survival risk in the evolution of LUAD.

\subsection{Prognostic Factors in Progressive Stages of LUAD}
\subsubsection{Data preprocessing and feature dimensionality reduction} 
Samples from each stage were classified to identify patient 
groups with differences in survival, but a significant portion 
of the follow-up data from clinical survival data was lost and 
there was no final patient survival time. 
To facilitate accurate identification of prognostic molecules 
of cancer without being disturbed by the missing clinical 
sample information, we only selected patient data with complete 
survival information for prognostic factors analysis 
(see Supplementary Table 2 for details). 
Considering that this kind of high-dimensional data still has 
many molecular characteristics even after DE analysis, those 
samples for commonly used ML classification 
algorithms are extremely insufficient. 
Therefore, for the purpose of finding prognostic factors, 
we used mutual information~\citep{KSG2004,R2014,KL1987} to screen 
out the features unrelated to the survival of patients. 
In the three LUAD stages of microRNA expression data, 57, 
66, and 51 molecules were screened to determine the 
survival probability of the patient.

\subsubsection{LightGBM models and SHAP algorithm}
\begin{figure*}[!ht]
\centering
\includegraphics[width=0.9\linewidth]{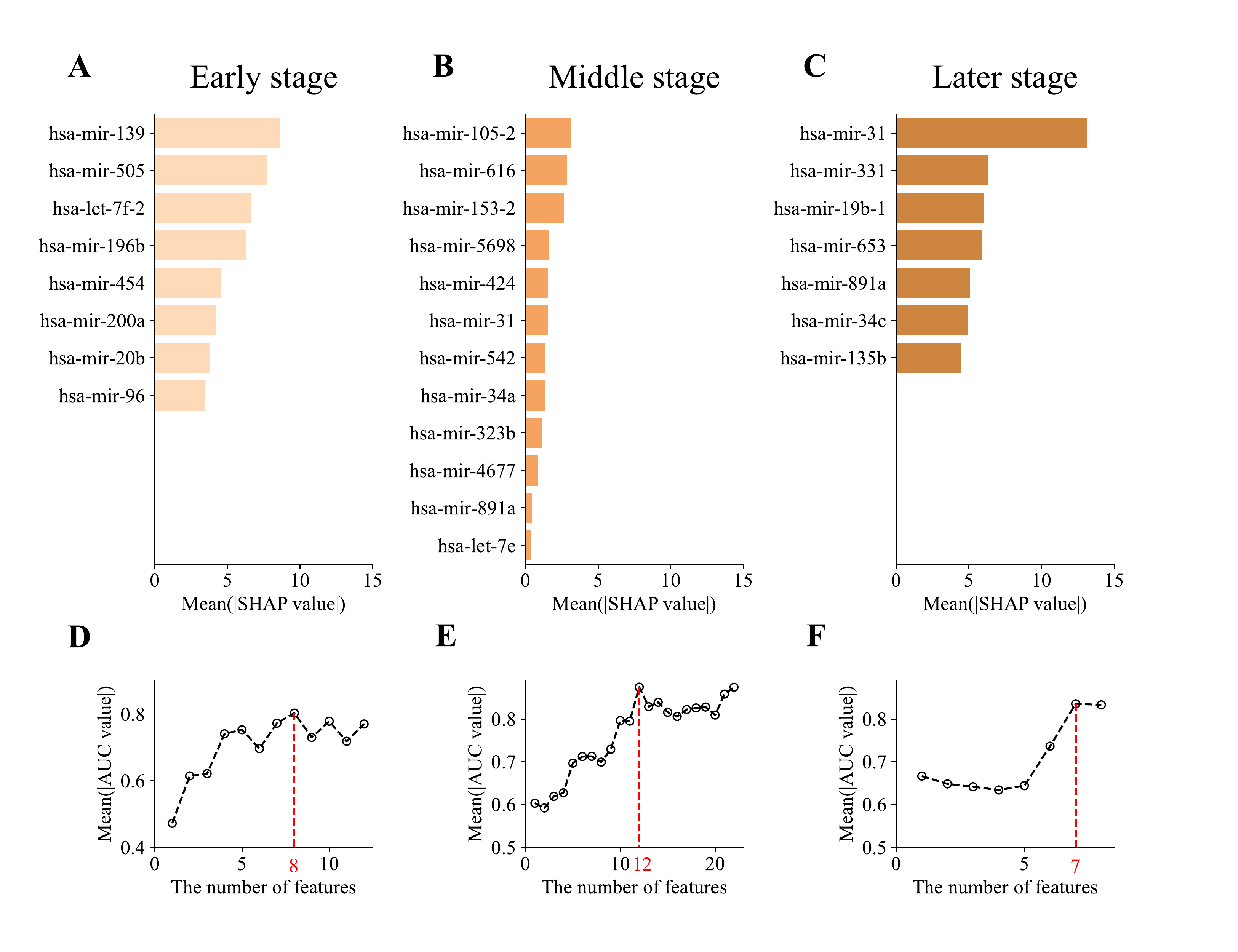}
\caption{\textbf{SHAP values of prognostic factors in three stages of LUAD}. 
The prognostic factors obtained after 3 rounds of classifier 
training and their corresponding contributions (SHAP value) to the model 
were displayed from high to low in the A. Early, B. Middle, and C. Later stages 
of LUAD. The following line charts (D. Early, E. Middle, and F. Later.)
described the average AUC value of the classifier in the 10-fold 
cross-validation when the top k features were selected in the 
round 2 model. The red dotted line indicates that the model performs 
best when k is 8, 12, and 7 respectively.}
\label{fig:prog}
\end{figure*}

Prognostic factors are closely related to the survival time 
of patients. 
For the purpose of exploring which biomolecules can affect 
the prognosis with lung cancer, a ML algorithm 
LightGBM~\citep{KMFWC2017,Q2020,ASSKK2021,DDYHH2020} was 
applied to predict the survival time of patients. 
Based on the gene expression data of microRNA, we constructed 
binary classifiers of the survival time in the early, middle, 
and later stages of LUAD. 
To characterize the importance of various molecules in a 
specific classifier, the SHAP value~\citep{LEL2018,LECDP2020} 
was measured.
At the same time, to better estimate the performance of the model, 
we used the 10-fold cross-validation training method. In this research, 
ACC and AUC are both the average results of the test sets in 10-fold 
cross-validation, as shown in Figure~\ref{fig:prog}.

Specifically, we first trained the binary LightGBM prediction models 
based on microRNA expression data, with AUC values of 0.62, 0.68, 
and 0.68 in three stages of LUAD, as shown in Table 1.
{\begin{table} [!ht]
\caption{The performance of 10-fold cross-validation of LightGBM classifier in LUAD three stges.}
\centering
\begin{tabular}{l c c c c}
\hline\hline
&{Round}&{Num. of factors}&{ACC}&{AUC} \\
\hline
\multirow{3}{*}{Early stage}&{1} &{57} &{0.70}&{0.62} \\
&{2} &{12} &{0.83}&{0.77} \\
&{3} &{8} &{0.86}&{0.81} \\
\hline
\multirow{3}{*}{Middle stage}&{1} &{66} &{0.71}&{0.68} \\
&{2} &{22} &{0.89}&{0.87} \\
&{3} &{12} &{0.88}&{0.88} \\
\hline
\multirow{3}{*}{Later stage}&{1} &{51} &{0.67}&{0.68} \\
&{2} &{8} &{0.82}&{0.83} \\
&{3} &{7} &{0.85}&{0.87} \\
\hline\hline
\end{tabular}
\end{table}
}

In order to further optimize the model and improve its performance, 
we calculated the contribution of each molecule to the model 
accuracy by SHAP algorithm and selected the top k microRNAs 
for a new round of classifiers training. 
Then, by comparing the performance of different classifiers, we 
acquired 12, 22, and 8 molecules (AUC: 0.77, 0.87, and 0.83) 
in the sequential three stages of LUAD, which were significantly 
higher than the performance of the previous model. 
In this round, the new classification models were evaluated, 
and the top 8, 12, and 7 molecules were again filtered out according 
to the SHAP value. 
Finally, the AUC of the classification models reached 0.81, 0.88, 
and 0.87 respectively. And when continuing to optimize the 
model with the interpretable methods, we found the classifier 
performance couldn't be improved by further reducing the number 
of microRNAs. So, we obtained three binary classifiers and got 
the microRNA molecules which are the key factors affecting 
the survival time of patients and potentially used to measure 
the prognosis according to the microRNA expression data with 
early, middle and later LUAD.

\subsubsection{Relationship between algorithm indexes and 
characteristics of gene expression}
It is worth mentioning that in the second round of classifier models, 
we carefully analyzed the gene expression characteristics of 
those molecules, and explored the bioinformatics significance 
of this ML method.
These LUAD-related top prognostic factors obtained from the binary 
classification model are independent variables from their expression 
characteristics, such as microRNA expression level or 
FC value Supplementary Figure 3.

\subsection{Key Diagnostic and Prognostic Factors}
\subsubsection{The dynamical interaction of diagnosis and 
prognosis factors in evolutionary LUAD}
\begin{figure*}[!ht]
\centering
\includegraphics[width=0.9\linewidth]{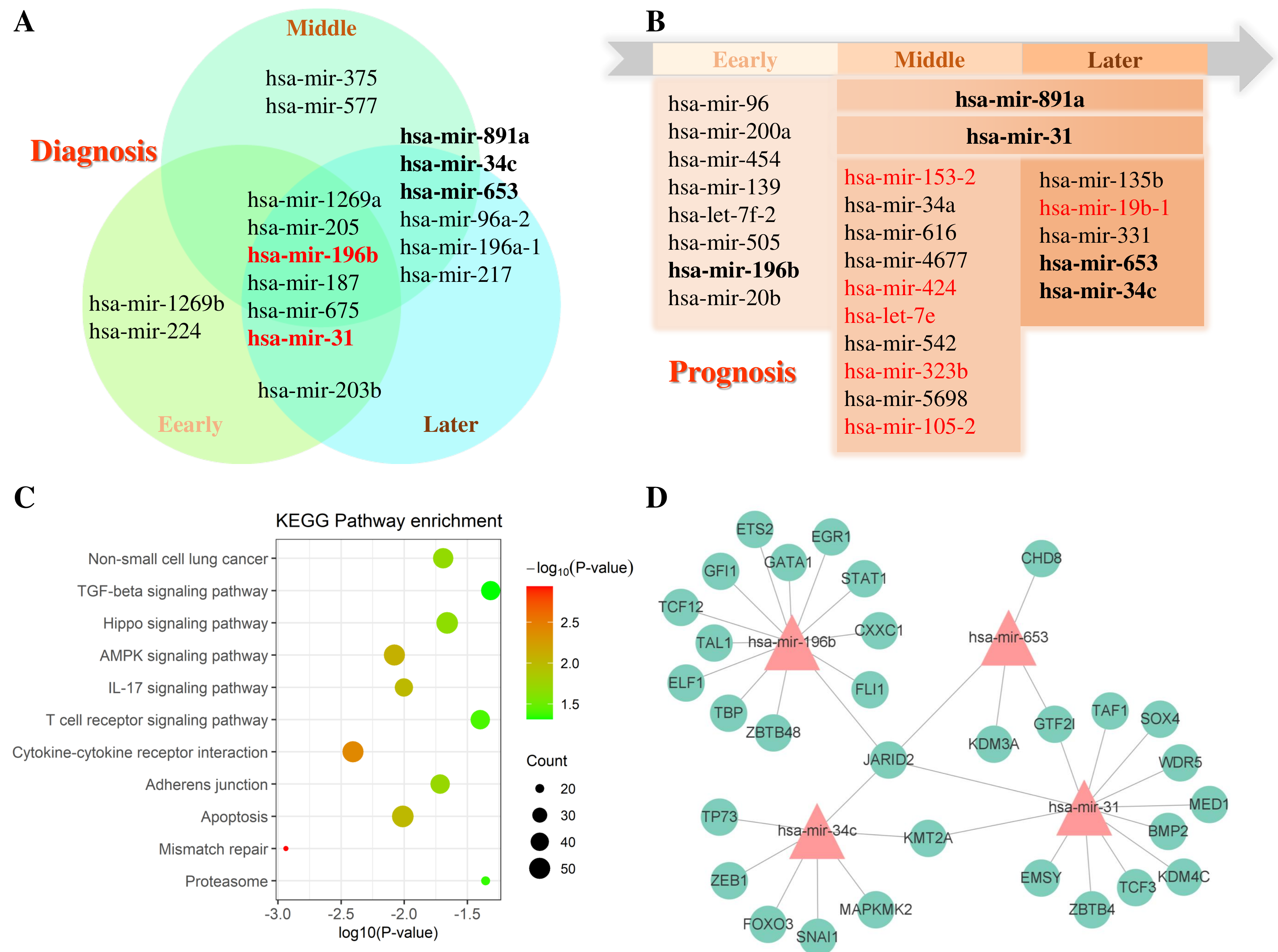}
\caption{\textbf{The key diagnostic and prognostic factors with the evolutionary LUAD}.
A. The diagnostic factors for LUAD stages were shown in a Venn diagram. 
B. An overall diagram of key prognostic factors in three stages of LUAD. 
The molecules in bold are both shared in diagnostic and prognostic factors,
and those in red have an effect on patient survival. 
The diagnostic factors in the dynamical LUAD have a very high-level overlap in various stages.
C. Those final microRNAs served as diagnostic factors in different LAUD stages were 
significantly enriched in some pivotal cancer-related KEGG pathways. 
D. The microRNA-TF interaction network, where the microRNAs belong to
both diagnostic and prognostic factors in our analyses.}	
\label{fig:markers}
\end{figure*}

We have developed two methods to find potential microRNAs 
which can be served as diagnostic and prognostic factors 
in the evolution of LUAD. 
Because there is no mechanism introduced in our methods 
process to exclude the distribution of biomarkers in various 
stages of cancer, usually there are some 
overlap biomarkers in different stages.
In Figure~\ref{fig:markers} A and B, the six microRNAs 
hsa-mir-1269a, hsa-mir-205, hsa-mir-196b, hsa-mir-187, 
hsa-mir-675, and hsa-mir-31 are used as diagnostic factors shared 
with the three stages of LUAD. However, there is no overlap 
of the prognostic factors in all LUAD stages.

In addition, the prognostic and diagnostic factors
relatively overlap in the middle and later stages in the 
evolution of the classification of LUAD samples. However, 
the early stage of TNM staging (Stage I) is indeed different 
from the later stage (stage III and IV).
To be specific, in the prognostic factors, hsa-mir-891a and 
hsa-mir-31 exist in both middle- and later-stage, while no 
molecule is the same in the early stage as that in the 
remaining two stages. For the biomarkers of different stages, 
between the middle- and later-stage, it contains many common 
biomarkers, such as hsa-mir-891a, hsa-mir-34c, hsa-mir-653, 
hsa-mir-196a-2, hsa-mir-196a-1, hsa-mir-217. In contrast, there 
is only the intersection of hsa-mir-203b in the early- and the 
later-stage, and even no common molecule with middle stage. 

\begin{figure*}[!ht]
\centering
\includegraphics[width=0.9\linewidth]{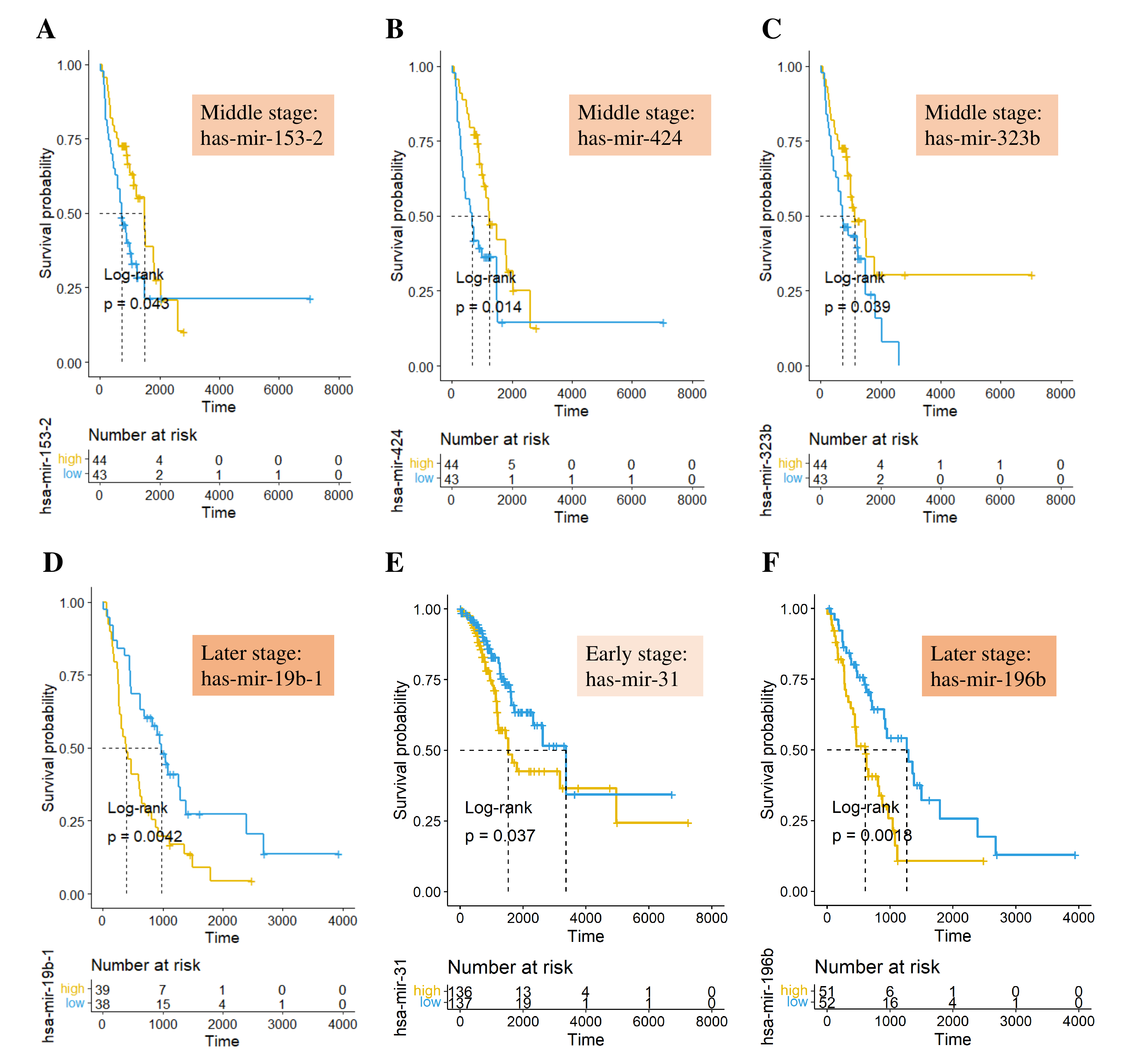}
\caption{\textbf{MicroRNAs that affect patient survival}.
The Kaplan-Meier analysis showed that the key microRNAs (four of prognostic 
factors and two of diagnostic factors) have a significant influence on patient 
survival in a specific LUAD stage.
Log-rank tests were used to analyze the Kaplan-Meier survival curve.}
\label{fig:sur}
\end{figure*}

It's worth noting that the five microRNAs of hsa-mir-196b, 
hsa-mir-31, hsa-mir-891a, hsa-mir-34c, and hsa-mir-653 in 
the three stages of LUAD evolution are not only used as 
diagnostic factors but also as prognostic factors. 
Among them, hsa-mir-31 and hsa-mir-196b as diagnosis 
of all stages of LUAD can influence the survival of patients 
in the early- and later-stage Figure~\ref{fig:sur}E and F. 

\subsubsection{Molecular heterogeneity of Stage II}
Generally, since changes of TNM pathological characteristics 
are continuous during deteriorative cancer, it cannot well 
distinguish the biomarkers in different stages of LUAD by the 
analysis of the gene expression data of microRNA transcripts.
This is consistent with the conclusion of the difference between the 
clinical classification and biomolecular characteristics of the
cancer stage. In the meantime, we also found that the 
heterogeneity of the second stage may be relatively high, this 
phenomenon is also consistent with the previous study
~\citep{WTLKH:2019,RBFS:2021}.
Detailed, in our research, the number of prognisitic microRNAs 
in stage II is larger and their SHAP value are relatively
smaller compared with other stages. Besides, the variance of 
eigenvector clustering height in the middle stage of LUAD 
(see Figure~\ref{fig:cluster}B) is larger which indicates a high 
degree of similarity within the clustering groups. 

\subsubsection{Functional validation of precursor microRNAs}
What is the function of those stem-loop microRNAs mined 
from lung cancer? To address this question, we a performed 
KEGG pathway enrichment analysis of mature microRNA 
corresponding to the precursor microRNA in our LUAD data. 
(Supplementary File 1 presents more details of the 
enrichment analysis of those precursor microRNAs.) 
The result displays that the biomarkers of microRNAs are 
strongly correlated with cancer-related pathways, 
such as Non-small cell lung cancer ($-\log_{10}{P} > 20$).
Besides, it is also mainly related to signal transduction 
and cellular immune functions, such as hsa-mir-196b, 
hsa-mir-31, hsa-mir-34c, hsa-mir-891a, etc. being involved 
in ``Non-small cell lung cancer,'' ``TGF-beta signaling pathway,'' 
``Hippo signaling pathway,'' and ``AMPK signaling pathway''. 
In addition, these microRNAs also work with ``IL-17 signaling pathway,'' 
``T cell receptor signaling pathway,'' ``Cytokine-cytokine 
receptor interaction,'' and other immune pathways are related, 
and may also be related to ``Apoptosis,'' ``Adherens junction,'' 
``Mismatch repair,'' and ``Proteasome biological process''. 

Taken together, the enrichment analysis of microRNAs 
reveals that our reconstructed diagnosis and prognosis 
factors exhibit a close correspondence to the development 
of LUAD and deterioration of physiological indicators 
from early- to later-stage, validating our systematic 
analysis method.

\subsubsection{MicroRNA targeted TF network} 
A very important part of the regulatory mechanism of 
microRNA is to regulate the transcription process by 
forming a complex with the protein-encoding mRNA
or protein that participates in the DNA trans regulation~\citep{WTLKH:2019}. 
Therefore, we had also investigated the interaction 
between microRNAs and transcription factors (TF) and 
found that some molecules are closely related to 
cancer, which shed light on the regulatory mechanism 
of a microRNA to control cancer progression, as shown in 
Figure~\ref{fig:markers}D. 

JARID2 encodes a Jumonji- and AT-rich interaction domain 
(ARID)-domain-containing protein. The encoded protein is 
a DNA-binding protein that functions as a transcriptional 
repressor. 
It regulates gene expression, but the precursor RNA mainly 
accumulates in the nucleus, which can also be combined with 
transcription factors to regulate the process of gene expression. 
This gene functions as a node gene for hsa-mir-196b, 
hsa-mir-31, hsa-mir-34c, and hsa-mir-653. 
We noticed that hsa-mir-196b interacts with EGR1, which 
is closely related to cancer~\citep{LFS2018,HXZYC2020,LMHZY2020}, 
and hsa-mir-31 is involved in the key gene SOX4 which may 
function in the apoptosis pathway leading to cell death 
as well as to tumorigenesis~\citep{HAVA2020,HSJZYJ2016,EBMMD2016}.

And hsa-mir-34c is related to the FOXO3 gene, which belongs 
to the forkhead family of TFs and likely functions as a 
trigger for apoptosis through the expression of genes necessary 
for cell death~\citep{HYLLM2020,LRT2014}.
Both hsa-mir-31 and hsa-mir-34c can target KMT2A protein. 
This gene encodes a transcriptional coactivator 
that plays an essential role in regulating gene expression 
during early development and hematopoiesis~\citep{SDAR2010}. 
The encoded protein contains conserved functional domains 
SET which is responsible for its histone H3 lysine 4 (H3K4) 
methyltransferase activity which mediates chromatin modifications 
associated with epigenetic transcriptional activation~\citep{SDAR2010}. 
The Gene Ontology (GO) annotations related to GTF2I (General 
Transcription Factor IIi) include DNA-binding transcription 
factor activity and mitogen-activated protein kinase binding. 
Among its related pathways are Assembly of RNA Polymerase-II 
Initiation Complex and Akt Signaling. Nevertheless, co-targeting 
of GTF2I from the precursor microRNA hsa-mir-31 and hsa-mir-653 
may be connected with cancer in our network. 
Supplementary File 2 and 3 present more details of the 
possible pathological associations of these key microRNAs 
as diagnostic and prognostic factors dated from the knowledge 
database~\citep{XDHW2013,HSGCZ2019}.

\subsubsection{MicroRNAs with a significant influence on patient survival}
Combining the microRNA gene expression data and the 
survival data of the samples during the evolution of 
lung cancer, we performed the K-M survival analysis
~\citep{KM:1958,Kaplan:1983} to study the relationship 
between these molecules and patient survival. 
The results showed that hsa-mir-153-2 (P-value, 0.043), 
hsa-mir-424 (P-value, 0.014), hsa-mir-323b (P-value, 0.039) 
significantly affects the survival of the middle stage 
samples Figure~\ref{fig:sur}, while hsa-mir-19b-1 
(P-value, 0.0042) has a significant effect on survival 
of later samples. Figure~\ref{fig:sur} A-C.
hsa-mir-31 (P-value, 0.0037) and hsa-mir-196b (P-value, 0.0018), 
as all-stage diagnostic factors, can affect the survival rate 
of patients in the early- and later-stage respectively 
Figure~\ref{fig:sur} E and F. 
In addition, several microRNAs affect the survival for 
all LUAD patients as shown in Supplementary Note 3 
and Supplementary Figure 4.

\section{Discussion}
LUAD has more pronounced genomic variations~\citep{Vergoulisetal:2011} 
than other lung cancer subtypes, which are rarely caused by 
a few genetic changes~\citep{Hsuetal:2014}, rendering necessary 
articulating alternative methodologies in order to obtain a 
reasonable understanding of the complicated mechanisms behind 
the evolution of cancer. 
In the past decades, it was widely believed 
that non-coding RNA had been involved in the regulation 
of gene expression and cancer occurrence and development.
Analyses based on the transcriptome data of non-coding 
RNA provide a promising approach to understanding microRNA 
regulation at the post-transcriptional level
~\citep{XZCKGL:2008,TKP:2014,AKS:2014}.
Although some work studied how multiple types of molecular 
interactions affect cancer, such as CeRNA-related 
studies~\citep{YYKCJ2019,ZFT2021} that match the relationship 
between endogenous competing RNAs to analyze biological functions and 
pathways closely related to lung cancer. 
Nevertheless, these studies ignore the tumor microenvironment 
and the accuracy of ceRNAs targeting genes. 

Therefore, from a clinical pathology view, the effects of microRNA 
on the stage characteristics and prognosis
of cancer patients were still not well understood.
In order to simplify the problem without introducing certain 
biological hypotheses such as the CeRNA hypothesis
~\citep{ENS:2007,SPTKP:2011,TRP:2014}, but based on 
the original transcriptome data. 
We hoped to comprehensively assess which molecules in the 
evolution of cancer are related to the patient's 
diagnosis and prognosis and how these key microRNAs 
evolve along with the lung cancer worsening. 
We systematically explored the relationship between the two 
functions of microRNA as the diagnostic and prognostic 
factors. Because of the limited sample cases and the high degree 
of molecular heterogeneity of cancer, it is infeasible to 
determine which methods are more suitable for finding 
biomarkers of LUAD in the process of microRNA transcriptional 
data. 
In our research, through a step-by-step method, combined with 
corresponding biological information and clinical data, 
we analyzed the microRNA expression data of LUAD patients in the 
early, middle, and later stages. 
Our analysis delineated the dynamical changes of 
these key microRNAs in the progression stages from 
early to later.
Beyond that, high molecular heterogeneity was also observed 
in the middle stage.

By performing KEGG pathway enrichment analysis on mature 
microRNAs linked by precursor, we had identified those key 
underpinning RNA molecules that are not only directly involved 
in various cancer pathways but also signal transduction 
as well as immune-related pathway (Figure~\ref{fig:markers}C). 
This is a strong indication that these microRNAs are involved 
in LUAD and its evolution.
Especially, by targeting TF, we found that some specific 
oncogene in the microRNA and TF network (Figure~\ref{fig:markers}D). 
Through the Kaplan-Meier survival analysis~\citep{KM:1958,Kaplan:1983}, 
we had also identified some key microRNAs that have a significant 
influence on the survival rate of LUAD patients (Figure~\ref{fig:sur}).
In a word, beyond the ``static'' information provided 
by the conventional gene ontology analysis of the four 
stages of LUAD, our framework of data process gives rise to a 
dynamic scenario for tumor progression as the TNM stage deteriorates. 

In particular, five microRNAs markedly acted as the diagnostic factors 
of stage classification and prognostic factors of LUAD patients. 
The first is hsa-mir-196b reported in recent studies, which 
promotes lung cancer cell migration and invasion through 
the targeting of GATA6~\citep{LFS2018}.
Another microRNA is hsa-mir-31, which is involved in the 
inhibition of specific tumor suppressants in human lung 
cancer~\citep{KS:2011}, and some important signaling 
pathways about lung cancer~\citep{EBMMD2016, HSJZYJ2016}. 
The third one is hsa-mir-34c, which participates 
in promoting invasion and migration of non-small cell 
lung cancer (NSCLC) by upregulating integrin alpha 2 
beta 1~\citep{HYLLM2020}, and its homolog may be served 
as a therapeutic target in human cancer~\citep{LRT2014}.
Nevertheless, hsa-mir-891a and hsa-mir-653 have not been 
widely reported.

The stem-loop microRNA we studied, which is the precursor 
of mature microRNA, can be regarded as a reference of mature 
microRNA helping to better understand the potential molecules 
of prognostic and diagnostic factors in the progression of LUAD.
This mechanism of action can be used for further research 
on potential drug targets.
The quantitative analysis of microRNA expression data is 
indeed significant for cancer development, validating and 
demonstrating the power of our framework of dynamical 
biomarkers research.

Taken together, based on the ncRNA transcriptome, we had 
developed two frameworks to analyze the processes of identifying 
microRNA function in LUAD patients.
Our analysis had yielded some diagnostic and prognostic factors closely 
related to the three evolutionary stages of LUAD. 
The comparative study of these biomarkers provides new insights 
into understanding cancer mechanisms and identifying targets for 
further drug development.

\section{Materials and Methods}
\subsection{Clinical and microRNA expression data of LUAD}
A total of 555 samples with the corresponding microRNA
expression profile data from The Cancer Genome Atlas (TCGA) were used.
We followed the Tumor-Node-Metastasis (TNM) staging criteria for malignant
tumors to classify LUAD into four stages. In particular, TNM is a standardized
classification system established by the International Association for the
Study of Lung Cancer to describe the development of lung cancer in terms of
size and spread, where ``T'' describes the size of the tumor and any
spread of cancer into nearby tissues, ``N'' denotes the spread of 
cancer to nearby lymph nodes, and ``M'' stands for metastasis, i.e., the
spread of cancer to other parts of the body~\citep{GPJKD:2007}. 
Detailed in Supplementary Table 4. For our datasets, we integrated the 
TNM stage III and IV as the later stage. 
Finally, We had that, of the remaining 542 samples, 273, 120, and 103
are labeled as early, middle, and later stages of LUAD, correspondingly, 
and 46 are normal samples in diagnosis research. 
In prognosis analysis, it requires pre-processing samples to match the 
clinical information and especially with not censored survival time data.
Thus, the numbers of LUAD samples are 126, 87, and 77 in three 
stages respectively. (More details in Supplementary Note 1)

\subsection{Differential Expression analysis}
Fold Change (FC) characterizes the relative expression level of samples of interest 
to that of the control samples. The individual RNA expression can 
be seen from the volcano map. 
We used the R-language ``Limma'' package to analyze the differentially 
expressed (DE) microRNA profile data in the three stages of LUAD, 
with the threshold of $\log_{2}{\mbox{FC}}$ absolute value for filtering 
the microRNAs set as 1, while ensuring that their P-value 
(t-statistic~\citep{phipson2016robust}, see Supplementary Note 4 for detail) 
is less than 0.05. 
So, RNAs with FC greater than 1 correspond to up-regulated genes, while 
those less than minus 1 to down-regulated genes. 
For the RNA expression data, the total number of stem-loop microRNAs is 1881. 
Finally, we then obtained that the numbers in the three stages of 
LUAD are 127, 130, and 131 respectively, as listed in Supplementary Table 1.

\subsection{KEGG enrichment analysis}
Gene enrichment analysis is a widely used approach to identify biological 
connections. We performed KEGG enrichment analysis for microRNAs of the LUAD,
taking into account the various biological processes 
in Figures~\ref{fig:DE}C and \ref{fig:markers}C.
In particular, the P-value determines whether any term annotates a specified
list of genes at a frequency greater than that which can be expected by chance,
as determined by the hypergeometric distribution:
\begin{equation} \label{eq:hypergeometric}
\begin{split}
p=1-\sum\limits_{i=0}^{k-1}\frac{{\binom{N-M}{n-i}}{\binom{M}{i}}}{\binom{N}{n}},
\end{split}
\end{equation}
where $N$ is the total number of genes in the background distribution, $M$ is
the number of genes within that distribution that are annotated (either
directly or indirectly) to the node of interest, $n$ is the size of the list
of genes of interest, and $k$ is the number of genes within that list which
are annotated to the node. The background distribution by default is all the
genes that have an annotation.

\subsection{Eigenvalue decomposition of microRNA expression data}
LUAD sample-RNA expression matrix ~\textbf{X}, the row $m$ is the number 
of different LUAD samples, and the column $n$ is the number of microRNAs.
\begin{equation}
	\mathbf{X} = \left(
   \begin{array}{cccc}
   x_{11} & x_{12} & \ldots & x_{1n}\\
   x_{21} & x_{22} & \ldots & x_{2n}\\
   \vdots & \vdots & \ddots & \vdots\\
   x_{m1} & x_{n2} & \ldots & x_{mn}\\
   \end{array} \right)\text{.}
   \end{equation}
Decentralize~\textbf{X} through minus the average expression value 
and calculate the Covariance matrix~\textbf{C}:
\begin{equation}
\mathbf{C}= \frac{\mathbf{X}^{T} \star \mathbf{X}} {n-1}\text{.}
\label{eq:pca_cov}
\end{equation}
We performed eigenvalue decomposition of matrix~\textbf{C} to obtain 
eigenvectors that correspond to the projection direction of the 
reconstructed data in the high-dimensional space. 
A load of eigenvector corresponds to each microRNA, so these eigenvectors 
are the preservation of microRNA gene expression characteristic information.
The degree of variation was calculated by dividing the variance of each 
eigenvalue by the sample size $n-1$.
Then we took the top $k$ eigenvectors whose degree of variation is greater than 
90$\%$ to obtain information that retains more than 90$\%$ of the original 
microRNA expression data.

\subsection{Survival curve analysis}
We used the Kaplan-Meier method~\citep{KM:1958,Kaplan:1983}, a non-parametric 
method, to estimate survival probability from the observed survival 
data~\citep{Smyth:2004}. The survival probability as a function 
of time is calculated according to 
\begin{equation}
	S(t_i) = S(t_{i-1})(1 - d_i/n_i),
\end{equation}
where $n_i$ is the number of patients who were alive before time 
$t_i$ and $d_i$ is the number of death events at $t_i$. The estimated 
probability $S(t_i)$ is a step function that changes value only at 
the time $i$ of each event.

Log-rank tests are applied to carry out univariate analysis 
of the Kaplan-Meier survival curve, which belongs to the 
chi-square test, where all time points of the sample survival 
information are equal (i.e., with weight setting to one).

\subsection{Mutual Information (MI)}
Mutual information is a powerful statistical method for feature 
selection in machine learning, which reduces the input size of 
the data set without affecting the most relevant features supporting 
classification or regression problems.
Feature selection can be performed by finding correlations between 
random features and mutual information (MI) between two discrete 
random features A = ($a_1$, $a_2$, …, $a_k$) and B = ($b_1$, $b_2$, …, $b_k$) 
is defined as
\begin{equation}
	I(A, B)=\sum_{a} \sum_{b} p(a, b) \log \frac{p(a, b)}{p(a) p(b)}.
\end{equation}

And when only two random variables are independent, MI equals zero, 
and a higher value means a higher dependence. 
In our research, we calculated the MI between each feature vector 
and the patient survival probability vector respectively and 
reduced the dimension of the data set by keeping the features with 
MI greater than 0. 

This method can greatly preserve the molecular information 
related to the prognosis targets. In addition, for the mutual 
information algorithm, we finally chose a relatively loose 
threshold to ensure that the potential predictive factors 
will not be discarded. 

\subsection{Light Gradient Boosting Machine (LightGBM)}
LightGBM is one of the ensemble modelings which is a method of 
creating strong learners by combining weak learners. 
It is a decision tree algorithm based on the histogram, which has 
the advantages of fast operation, high accuracy, strong robustness, 
and so on. It adopts Gradient-Based One-Side Sampling (GOSS) and 
pays more attention to the samples with insufficient training 
without excessively changing the distribution of the original 
data set, thereby reducing the error and improving the accuracy. 
At the same time, while ensuring high efficiency, the leaf-by-leaf 
algorithm with depth limitation is adopted, which avoids over-fitting. 
Based on the above advantages, LightGBM has been widely used in 
various machine learning tasks, and it has a very good performance 
in medical data tasks.

\subsection{SHAPley Additive explanation (SHAP)}
There are many ways to calculate the importance of the features. 
Among them, the SHAP value can be viewed through the ``Consistent 
Individualized Feature Attribution for Tree Ensembles''~\citep{LEL2018}, 
which has good consistency and accuracy in calculating 
feature importance.
Inspired by cooperative game theory, SHAP constructed an 
additive explanation model, and all characteristics were 
regarded as ``contributors''. For each prediction sample, 
the model will generate a prediction value, and the SHAP value 
is the assigned value of each feature in the sample.

Assuming that sample $i$ is $x_i$, the $j-th$ feature of sample 
$i$ is $x_{i j}$, the predicted value of the model for this 
sample is $y_i$, and the baseline of the whole model (usually 
the average value of the target variables for all samples) is 
ybase, and the SHAP value follows the following equation.
\begin{equation}
y_{i}=y_{\text {base }}+f\left(x_{i 1}\right)+f\left(x_{i 2}\right)+\ldots+f\left(x_{i k}\right)
\end{equation}
Where $f(x_{i j})$ is the SHAP value of $x_{i j}$. Intuitively, 
$f(x_{i j})$ is the contribution of the first feature in 
the $j-th$ sample to the final predicted value $y_{i}$. When 
$f(x_{i j})$ $>$0, it means that this feature improves the 
predicted value and has a positive effect; On the contrary, 
this feature will lower the predicted value and has a negative impact. 

\subsection{The framework of analysis}
\begin{figure*}[!ht]
\centering
\includegraphics[width=0.9\linewidth]{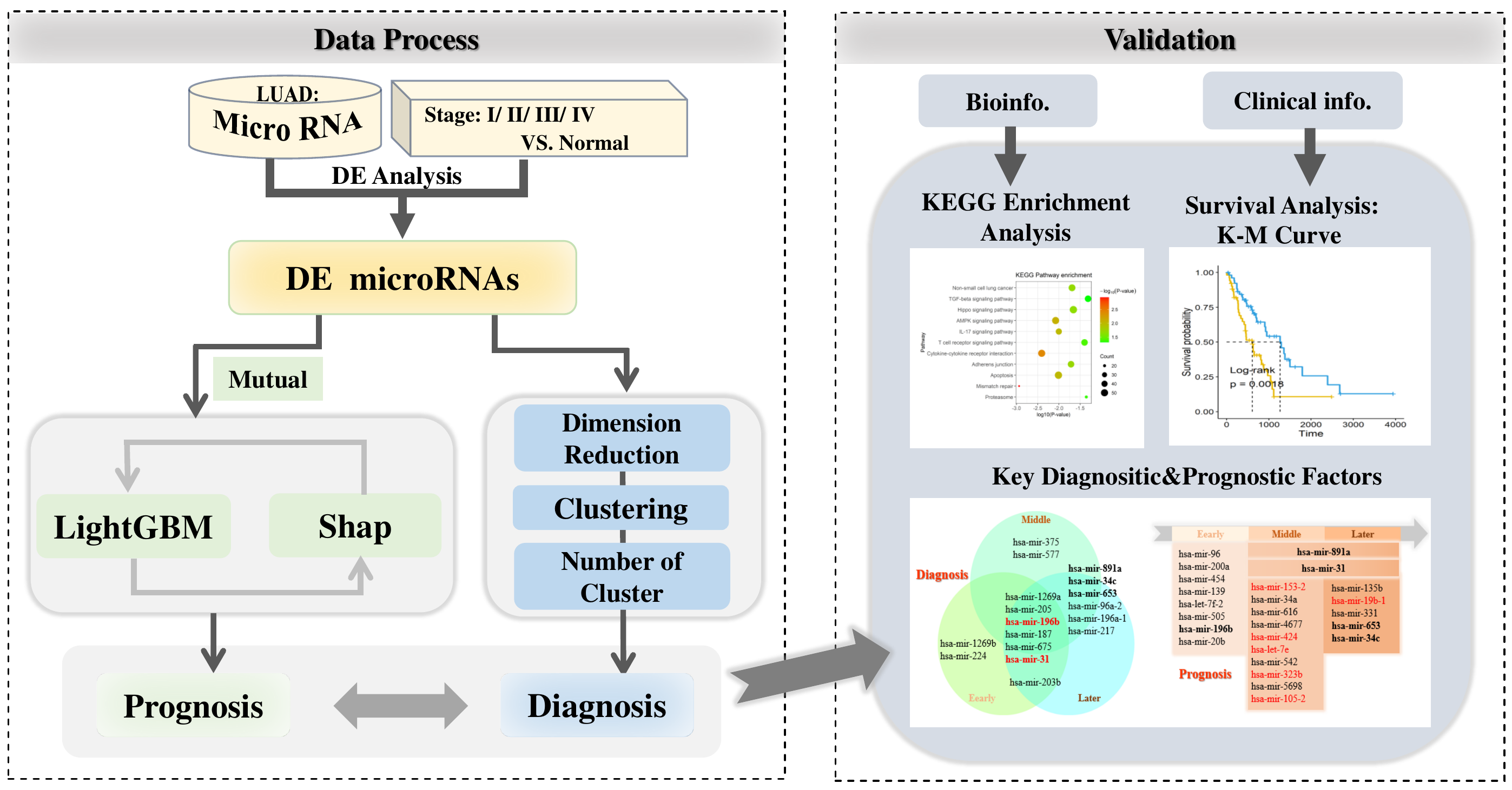}
\caption{\textbf{Overall workflow chart of proposed framework of dynamical biomarkers
analysis of lung cancer}. LUAD samples were divided into three stages according 
to the TNM criteria and analyzed by Differential Expression (DE) microRNAs.
Then data reduction, machine learning, and clustering were performed on the 
gene expression data to uncover the corresponding diagnostic and prognostic 
factors for each of the three LUAD stages. Statistic analyses about the 
indicators of algorithms and gene expression characteristics helped to verify 
these key biomarkers as independent molecular variables (see Supplementary Figures 2 and 3 ).
Finally, the microRNA targeted TF network, gene pathway enrichment, 
and survival analyses were collectively carried out
from bioinformatics and clinical data.}
\label{fig:flow}
\end{figure*}

Our articulated framework of dynamical analysis combines the
following methods: DE analysis, training classification ML 
for prognosis, searching for diagnostic factors, KEGG 
enrichment analysis, and survival analysis. A flow chart of 
these methods was illustrated in Figure~\ref{fig:flow}. 
In the section \textbf{Materials and Methods}, each of 
the methods were described.

\subsection{Computational packages and database}
Data processing and statistical analysis used R language 
(v.3.5.1) and Python (v.3.9). Analyze DE RNAs using the ``Limma''
package~\citep{ZWWYWWYS:2018}, and plot the heatmap of the enrichment analysis
and the Kaplan-Meier survival curves using the ``heatmap,'' ``survival,'' and
``survminer'' packages. 
In the search for prognostic factors, Python was the main language. 
``Pandas'' and ``Matplotlib'' were used for data analyzing 
and visualization; ``SKlearn'', ``LightGBM,'' and ``SHAP'' for the classifier.
The various microRNA-TF networks were visualized via Cytoscape (v3.6.1). 
Finally, The KEGG functional enrichment analysis used the online tool 
miEAA (2.0)~\citep{KFSSG2020}.

\section{Key points}
\begin{itemize}
\item    Through integrating clinical information into the transcripts expression data, 
we had developed two frameworks to analyze the processes of identifying microRNA clinical function: 
the unsupervised hierarchical clustering was used to find the diagnostic factors and a classification 
framework to screen out the prognostic factors.

\item Our analysis delineated the dynamical changes of these key microRNAs in the progression stages from early to later. 
Beyond that, high molecular heterogeneity was also observed in Stage II in LUAD.

\item We found that several microRNAs impact the survival risk of patients and some of them target important TF.

\item Five microRNAs (hsa-mir-196b, hsa-mir-31, hsa-mir-891a, hsa-mir-34c, and hsa-mir-653) can markedly serve as not only potential
diagnostic factors of stage classification but also prognostic tools in the monitoring of lung cancer.
\end{itemize}

\section{Competing interests}
There is NO Competing Interest.

\section{Author contributions statement}
Z.-T.B. and Q.-N.Z. designed research; D.K. and K.W. analyzed data and performed research; 
all authors wrote and reviewed the manuscript.

\section{Acknowledgments}
This work is supported by Energy Science and Technology Guangdong Laboratory (No.HND20TDZLZL00)

\bibliographystyle{abbrvnat}

\begin{thebibliography}{71}
	\providecommand{\natexlab}[1]{#1}
	\providecommand{\url}[1]{\texttt{#1}}
	\expandafter\ifx\csname urlstyle\endcsname\relax
	  \providecommand{\doi}[1]{doi: #1}\else
	  \providecommand{\doi}{doi: \begingroup \urlstyle{rm}\Url}\fi
	
	\bibitem[Aboutalebi et~al.(2020)Aboutalebi, Bahrami, Soleimani, Saeedi,
	  Rahmani, Khazaei, Fiuji, Shafiee, Ferns, Avan, et~al.]{ABSSR2020}
	H.~Aboutalebi, A.~Bahrami, A.~Soleimani, N.~Saeedi, F.~Rahmani, M.~Khazaei,
	  H.~Fiuji, M.~Shafiee, G.~A. Ferns, A.~Avan, et~al.
	\newblock The diagnostic, prognostic and therapeutic potential of circulating
	  micrornas in ovarian cancer.
	\newblock \emph{The international journal of biochemistry \& cell biology},
	  124:\penalty0 105765, 2020.
	
	\bibitem[Adams et~al.(2014)Adams, Kasinski, and Slack]{AKS:2014}
	B.~D. Adams, A.~L. Kasinski, and F.~J. Slack.
	\newblock {Aberrant regulation and function of microRNAs in cancer}.
	\newblock \emph{Curr. Biol.}, 24\penalty0 (16):\penalty0 R762--R776, 2014.
	
	\bibitem[Ahn et~al.(2021)Ahn, So, Synn, Kim, Kim, Byeon, Kim, Heo, Yang, Yun,
	  et~al.]{ASSKK2021}
	B.-C. Ahn, J.-W. So, C.-B. Synn, T.~H. Kim, J.~H. Kim, Y.~Byeon, Y.~S. Kim,
	  S.~G. Heo, S.-D. Yang, M.~R. Yun, et~al.
	\newblock Clinical decision support algorithm based on machine learning to
	  assess the clinical response to anti--programmed death-1 therapy in patients
	  with non--small-cell lung cancer.
	\newblock \emph{European Journal of Cancer}, 153:\penalty0 179--189, 2021.
	
	\bibitem[Anastasiadou et~al.(2018)Anastasiadou, Jacob, and Slack]{AJS:2018}
	E.~Anastasiadou, L.~S. Jacob, and F.~J. Slack.
	\newblock {Non-coding RNA networks in cancer}.
	\newblock \emph{Nat. Rev. Cancer}, 18\penalty0 (1):\penalty0 5, 2018.
	
	\bibitem[Asakura et~al.(2020)Asakura, Kadota, Matsuzaki, Yoshida, Yamamoto,
	  Nakagawa, Takizawa, Aoki, Nakamura, Miura, et~al.]{AKMYY2020}
	K.~Asakura, T.~Kadota, J.~Matsuzaki, Y.~Yoshida, Y.~Yamamoto, K.~Nakagawa,
	  S.~Takizawa, Y.~Aoki, E.~Nakamura, J.~Miura, et~al.
	\newblock A mirna-based diagnostic model predicts resectable lung cancer in
	  humans with high accuracy.
	\newblock \emph{Communications biology}, 3\penalty0 (1):\penalty0 1--9, 2020.
	
	\bibitem[Bartel(2018)]{B2018}
	D.~P. Bartel.
	\newblock Metazoan micrornas.
	\newblock \emph{Cell}, 173\penalty0 (1):\penalty0 20--51, 2018.
	
	\bibitem[Chen et~al.(2019)Chen, Heikkinen, Wang, Yang, Sun, and
	  Wong]{CHWYS2019}
	L.~Chen, L.~Heikkinen, C.~Wang, Y.~Yang, H.~Sun, and G.~Wong.
	\newblock Trends in the development of mirna bioinformatics tools.
	\newblock \emph{Briefings in bioinformatics}, 20\penalty0 (5):\penalty0
	  1836--1852, 2019.
	
	\bibitem[Dama et~al.(2021)Dama, Colangelo, Fina, Cremonesi, Kallikourdis,
	  Veronesi, and Bianchi]{DCFCK2021}
	E.~Dama, T.~Colangelo, E.~Fina, M.~Cremonesi, M.~Kallikourdis, G.~Veronesi, and
	  F.~Bianchi.
	\newblock Biomarkers and lung cancer early detection: State of the art.
	\newblock \emph{Cancers}, 13\penalty0 (15):\penalty0 3919, 2021.
	
	\bibitem[Dong et~al.(2020)Dong, Dan, Yawen, Haibo, Huan, Mengqi, Linglong, and
	  Zhao]{DDYHH2020}
	X.~Dong, X.~Dan, A.~Yawen, X.~Haibo, L.~Huan, T.~Mengqi, C.~Linglong, and
	  R.~Zhao.
	\newblock Identifying sarcopenia in advanced non-small cell lung cancer
	  patients using skeletal muscle ct radiomics and machine learning.
	\newblock \emph{Thoracic cancer}, 11\penalty0 (9):\penalty0 2650--2659, 2020.
	
	\bibitem[Ebert et~al.(2007)Ebert, Neilson, and Sharp]{ENS:2007}
	M.~S. Ebert, J.~R. Neilson, and P.~A. Sharp.
	\newblock {MicroRNA sponges: competitive inhibitors of small RNAs in mammalian
	  cells}.
	\newblock \emph{Nat. Meth.}, 4\penalty0 (9):\penalty0 721, 2007.
	
	\bibitem[Edmonds et~al.(2016)Edmonds, Boyd, Moyo, Mitra, Duszynski, Arrate,
	  Chen, Zhao, Blackwell, Andl, et~al.]{EBMMD2016}
	M.~D. Edmonds, K.~L. Boyd, T.~Moyo, R.~Mitra, R.~Duszynski, M.~P. Arrate,
	  X.~Chen, Z.~Zhao, T.~S. Blackwell, T.~Andl, et~al.
	\newblock Microrna-31 initiates lung tumorigenesis and promotes mutant
	  kras-driven lung cancer.
	\newblock \emph{The Journal of clinical investigation}, 126\penalty0
	  (1):\penalty0 349--364, 2016.
	
	\bibitem[Epsi et~al.(2019)Epsi, Panja, Pine, and Mitrofanova]{EPSI:2019}
	N.~J. Epsi, S.~Panja, S.~R. Pine, and A.~Mitrofanova.
	\newblock {pathCHEMO}, a generalizable computational framework uncovers
	  molecular pathways of chemoresistance in lung adenocarcinoma.
	\newblock \emph{Commun. Biol.}, 2\penalty0 (1):\penalty0 1--13, 2019.
	
	\bibitem[Farsi(2021)]{F2021}
	M.~Farsi.
	\newblock Filter-based feature selection and machine-learning classification of
	  cancer data.
	\newblock \emph{INTELLIGENT AUTOMATION AND SOFT COMPUTING}, 28\penalty0
	  (1):\penalty0 83--92, 2021.
	
	\bibitem[Gebert and MacRae(2019)]{GM2019}
	L.~F. Gebert and I.~J. MacRae.
	\newblock Regulation of microrna function in animals.
	\newblock \emph{Nature reviews Molecular cell biology}, 20\penalty0
	  (1):\penalty0 21--37, 2019.
	
	\bibitem[Goldstraw et~al.(2007)Goldstraw, Crowley, Chansky, Giroux, Groome,
	  Rami-Porta, Postmus, Rusch, Sobin, for the Study~of Lung Cancer International
	  Staging~Committee, et~al.]{GPJKD:2007}
	P.~Goldstraw, J.~Crowley, K.~Chansky, D.~J. Giroux, P.~A. Groome,
	  R.~Rami-Porta, P.~E. Postmus, V.~Rusch, L.~Sobin, I.~A. for the Study~of Lung
	  Cancer International Staging~Committee, et~al.
	\newblock {The IASLC Lung Cancer Staging Project: proposals for the revision of
	  the TNM stage groupings in the forthcoming (seventh) edition of the TNM
	  Classification of malignant tumours}.
	\newblock \emph{J. Thoracic Oncol.}, 2\penalty0 (8):\penalty0 706--714, 2007.
	
	\bibitem[Guinde et~al.(2018)Guinde, Frankel, Perrin, Delecourt, L{\'e}vy,
	  Barlesi, Astoul, Roll, and Kaspi]{Guindeetal:2018}
	J.~Guinde, D.~Frankel, S.~Perrin, V.~Delecourt, N.~L{\'e}vy, F.~Barlesi,
	  P.~Astoul, P.~Roll, and E.~Kaspi.
	\newblock Lamins in lung cancer: Biomarkers and key factors for disease
	  progression through {miR-9} regulation?
	\newblock \emph{Cell}, 7\penalty0 (7):\penalty0 78, 2018.
	
	\bibitem[Hanieh et~al.(2020)Hanieh, Ahmed, Vishnubalaji, and Alajez]{HAVA2020}
	H.~Hanieh, E.~A. Ahmed, R.~Vishnubalaji, and N.~M. Alajez.
	\newblock Sox4: epigenetic regulation and role in tumorigenesis.
	\newblock In \emph{Seminars in Cancer Biology}, volume~67, pages 91--104.
	  Elsevier, 2020.
	
	\bibitem[Hong et~al.(2020)Hong, Chuang, Huang, Weng, Chen, Chang, Liao, and
	  Huang]{HCHWC2020}
	H.-C. Hong, C.-H. Chuang, W.-C. Huang, S.-L. Weng, C.-H. Chen, K.-H. Chang,
	  K.-W. Liao, and H.-D. Huang.
	\newblock A panel of eight micrornas is a good predictive parameter for
	  triple-negative breast cancer relapse.
	\newblock \emph{Theranostics}, 10\penalty0 (19):\penalty0 8771, 2020.
	
	\bibitem[Hou et~al.(2016)Hou, Sun, Jiang, Zheng, Yang, Ji, Liang, Shi, Zhang,
	  Liu, et~al.]{HSJZYJ2016}
	C.~Hou, B.~Sun, Y.~Jiang, J.~Zheng, N.~Yang, C.~Ji, Z.~Liang, J.~Shi, R.~Zhang,
	  Y.~Liu, et~al.
	\newblock Microrna-31 inhibits lung adenocarcinoma stem-like cells via
	  down-regulation of met-pi3k-akt signaling pathway.
	\newblock \emph{Anti-Cancer Agents in Medicinal Chemistry (Formerly Current
	  Medicinal Chemistry-Anti-Cancer Agents)}, 16\penalty0 (4):\penalty0 501--518,
	  2016.
	
	\bibitem[Hsu et~al.(2014)Hsu, Tseng, Shrestha, Lin, Khaleel, Chou, Chu, Huang,
	  Lin, Ho, et~al.]{Hsuetal:2014}
	S.-D. Hsu, Y.-T. Tseng, S.~Shrestha, Y.-L. Lin, A.~Khaleel, C.-H. Chou, C.-F.
	  Chu, H.-Y. Huang, C.-M. Lin, S.-Y. Ho, et~al.
	\newblock {miRTarBase update 2014: an information resource for experimentally
	  validated miRNA-target interactions}.
	\newblock \emph{Nucl. Acids Res.}, 42\penalty0 (D1):\penalty0 D78--D85, 2014.
	
	\bibitem[Huang et~al.(2020{\natexlab{a}})Huang, Yan, Liu, Lin, Ma, Zhang, Dai,
	  Li, Guo, Chen, et~al.]{HYLLM2020}
	W.~Huang, Y.~Yan, Y.~Liu, M.~Lin, J.~Ma, W.~Zhang, J.~Dai, J.~Li, Q.~Guo,
	  H.~Chen, et~al.
	\newblock Exosomes with low mir-34c-3p expression promote invasion and
	  migration of non-small cell lung cancer by upregulating integrin
	  $\alpha$2$\beta$1.
	\newblock \emph{Signal transduction and targeted therapy}, 5\penalty0
	  (1):\penalty0 1--13, 2020{\natexlab{a}}.
	
	\bibitem[Huang et~al.(2020{\natexlab{b}})Huang, Xiao, Zhu, Yu, Cao, Zhang, Li,
	  Zhu, Wu, Zheng, et~al.]{HXZYC2020}
	X.~Huang, S.~Xiao, X.~Zhu, Y.~Yu, M.~Cao, X.~Zhang, S.~Li, W.~Zhu, F.~Wu,
	  X.~Zheng, et~al.
	\newblock mir-196b-5p-mediated downregulation of fas promotes nsclc progression
	  by activating il6-stat3 signaling.
	\newblock \emph{Cell death \& disease}, 11\penalty0 (9):\penalty0 1--13,
	  2020{\natexlab{b}}.
	
	\bibitem[Huang et~al.(2019)Huang, Shi, Gao, Cui, Zhang, Li, Zhou, and
	  Cui]{HSGCZ2019}
	Z.~Huang, J.~Shi, Y.~Gao, C.~Cui, S.~Zhang, J.~Li, Y.~Zhou, and Q.~Cui.
	\newblock Hmdd v3. 0: a database for experimentally supported human
	  microrna--disease associations.
	\newblock \emph{Nucleic acids research}, 47\penalty0 (D1):\penalty0
	  D1013--D1017, 2019.
	
	\bibitem[Kaplan(1983)]{Kaplan:1983}
	E.~L. Kaplan.
	\newblock This week's citation classic.
	\newblock \emph{Curr. Cont.}, 24:\penalty0 14, 1983.
	
	\bibitem[Kaplan and Meier(1958)]{KM:1958}
	E.~L. Kaplan and P.~Meier.
	\newblock Nonparametric estimation from incomplete observations.
	\newblock \emph{J. Ame. Stat. Asso.}, 53\penalty0 (282):\penalty0 457--481,
	  1958.
	
	\bibitem[Kasinski and Slack(2011)]{KS:2011}
	A.~L. Kasinski and F.~J. Slack.
	\newblock {MicroRNAs en route to the clinic: progress in validating and
	  targeting microRNAs for cancer therapy}.
	\newblock \emph{Nat. Rev. Cancer}, 11\penalty0 (12):\penalty0 849, 2011.
	
	\bibitem[Ke et~al.(2017)Ke, Meng, Finley, Wang, Chen, Ma, Ye, and
	  Liu]{KMFWC2017}
	G.~Ke, Q.~Meng, T.~Finley, T.~Wang, W.~Chen, W.~Ma, Q.~Ye, and T.-Y. Liu.
	\newblock Lightgbm: A highly efficient gradient boosting decision tree.
	\newblock \emph{Advances in neural information processing systems},
	  30:\penalty0 3146--3154, 2017.
	
	\bibitem[Kern et~al.(2020)Kern, Fehlmann, Solomon, Schwed, Grammes, Backes, Van
	  Keuren-Jensen, Craig, Meese, and Keller]{KFSSG2020}
	F.~Kern, T.~Fehlmann, J.~Solomon, L.~Schwed, N.~Grammes, C.~Backes, K.~Van
	  Keuren-Jensen, D.~W. Craig, E.~Meese, and A.~Keller.
	\newblock mieaa 2.0: integrating multi-species microrna enrichment analysis and
	  workflow management systems.
	\newblock \emph{Nucleic acids research}, 48\penalty0 (W1):\penalty0 W521--W528,
	  2020.
	
	\bibitem[Kozachenko and Leonenko(1987)]{KL1987}
	L.~Kozachenko and N.~N. Leonenko.
	\newblock Sample estimate of the entropy of a random vector.
	\newblock \emph{Problemy Peredachi Informatsii}, 23\penalty0 (2):\penalty0
	  9--16, 1987.
	
	\bibitem[Kraskov et~al.(2004)Kraskov, St{\"o}gbauer, and Grassberger]{KSG2004}
	A.~Kraskov, H.~St{\"o}gbauer, and P.~Grassberger.
	\newblock Estimating mutual information.
	\newblock \emph{Physical review E}, 69\penalty0 (6):\penalty0 066138, 2004.
	
	\bibitem[Li et~al.(2018)Li, Feng, and Shi]{LFS2018}
	H.~Li, C.~Feng, and S.~Shi.
	\newblock mir-196b promotes lung cancer cell migration and invasion through the
	  targeting of gata6.
	\newblock \emph{Oncology letters}, 16\penalty0 (1):\penalty0 247--252, 2018.
	
	\bibitem[Li et~al.(2014)Li, Ren, and Tang]{LRT2014}
	X.~Li, Z.~Ren, and J.~Tang.
	\newblock Microrna-34a: a potential therapeutic target in human cancer.
	\newblock \emph{Cell death \& disease}, 5\penalty0 (7):\penalty0 e1327--e1327,
	  2014.
	
	\bibitem[Liang et~al.(2020)Liang, Meng, Huang, Zhu, Yin, Wang, Fassan, Yu,
	  Kudo, Xiao, et~al.]{LMHZY2020}
	G.~Liang, W.~Meng, X.~Huang, W.~Zhu, C.~Yin, C.~Wang, M.~Fassan, Y.~Yu,
	  M.~Kudo, S.~Xiao, et~al.
	\newblock mir-196b-5p--mediated downregulation of tspan12 and gata6 promotes
	  tumor progression in non-small cell lung cancer.
	\newblock \emph{Proceedings of the National Academy of Sciences}, 117\penalty0
	  (8):\penalty0 4347--4357, 2020.
	
	\bibitem[Liu et~al.(2010)Liu, Friggeri, Yang, Milosevic, Ding, Thannickal,
	  Kaminski, and Abraham]{Liuetal:2010a}
	G.~Liu, A.~Friggeri, Y.~Yang, J.~Milosevic, Q.~Ding, V.~J. Thannickal,
	  N.~Kaminski, and E.~Abraham.
	\newblock {miR-21 mediates fibrogenic activation of pulmonary fibroblasts and
	  lung fibrosis}.
	\newblock \emph{J. Exp. Med.}, 207\penalty0 (8):\penalty0 1589--1597, 2010.
	
	\bibitem[Liu et~al.(2016)Liu, Beyer, and Aebersold]{LBA2016}
	Y.~Liu, A.~Beyer, and R.~Aebersold.
	\newblock On the dependency of cellular protein levels on mrna abundance.
	\newblock \emph{Cell}, 165\penalty0 (3):\penalty0 535--550, 2016.
	
	\bibitem[Lopez-Rincon et~al.(2019)Lopez-Rincon, Martinez-Archundia,
	  Martinez-Ruiz, Schoenhuth, and Tonda]{LMMST2019}
	A.~Lopez-Rincon, M.~Martinez-Archundia, G.~U. Martinez-Ruiz, A.~Schoenhuth, and
	  A.~Tonda.
	\newblock Automatic discovery of 100-mirna signature for cancer classification
	  using ensemble feature selection.
	\newblock \emph{BMC bioinformatics}, 20\penalty0 (1):\penalty0 1--17, 2019.
	
	\bibitem[Lundberg et~al.(2018)Lundberg, Erion, and Lee]{LEL2018}
	S.~M. Lundberg, G.~G. Erion, and S.-I. Lee.
	\newblock Consistent individualized feature attribution for tree ensembles.
	\newblock \emph{arXiv preprint arXiv:1802.03888}, 2018.
	
	\bibitem[Lundberg et~al.(2020)Lundberg, Erion, Chen, DeGrave, Prutkin, Nair,
	  Katz, Himmelfarb, Bansal, and Lee]{LECDP2020}
	S.~M. Lundberg, G.~Erion, H.~Chen, A.~DeGrave, J.~M. Prutkin, B.~Nair, R.~Katz,
	  J.~Himmelfarb, N.~Bansal, and S.-I. Lee.
	\newblock From local explanations to global understanding with explainable ai
	  for trees.
	\newblock \emph{Nature machine intelligence}, 2\penalty0 (1):\penalty0 56--67,
	  2020.
	
	\bibitem[Luo et~al.(2019)Luo, Liao, and Zhu]{LLZ2019}
	S.-S. Luo, X.-W. Liao, and X.-D. Zhu.
	\newblock Genome-wide analysis to identify a novel microrna signature that
	  predicts survival in patients with stomach adenocarcinoma.
	\newblock \emph{Journal of Cancer}, 10\penalty0 (25):\penalty0 6298, 2019.
	
	\bibitem[Ma et~al.(2020)Ma, Geng, Meng, Yan, and Song]{MGMYS2020}
	B.~Ma, Y.~Geng, F.~Meng, G.~Yan, and F.~Song.
	\newblock Identification of a sixteen-gene prognostic biomarker for lung
	  adenocarcinoma using a machine learning method.
	\newblock \emph{Journal of Cancer}, 11\penalty0 (5):\penalty0 1288, 2020.
	
	\bibitem[Molina et~al.(2008)Molina, Yang, Cassivi, Schild, and
	  Adjei]{MYCSA:2008}
	J.~R. Molina, P.~Yang, S.~D. Cassivi, S.~E. Schild, and A.~A. Adjei.
	\newblock Non-small cell lung cancer: Epidemiology, risk factors, treatment,
	  and survivorship.
	\newblock \emph{Mayo Clin. Proc.}, 83\penalty0 (5):\penalty0 584--594, 2008.
	
	\bibitem[Naeli et~al.(2020)Naeli, Yousefi, Ghasemi, Savardashtaki, and
	  Mirzaei]{NYGSM:2020}
	P.~Naeli, F.~Yousefi, Y.~Ghasemi, A.~Savardashtaki, and H.~Mirzaei.
	\newblock The role of micrornas in lung cancer: implications for diagnosis and
	  therapy.
	\newblock \emph{Current molecular medicine}, 20\penalty0 (2):\penalty0 90--101,
	  2020.
	
	\bibitem[Network and Others(2014)]{CGARN:2014}
	C.~G. A.~R. Network and Others.
	\newblock Comprehensive molecular profiling of lung adenocarcinoma.
	\newblock \emph{Nature}, 511\penalty0 (7511):\penalty0 543, 2014.
	
	\bibitem[Pandey et~al.(2021)Pandey, Mukhopadhyay, Sharawat, and
	  Kumar]{PMSK:2021}
	M.~Pandey, A.~Mukhopadhyay, S.~K. Sharawat, and S.~Kumar.
	\newblock Role of micrornas in regulating cell proliferation, metastasis and
	  chemoresistance and their applications as cancer biomarkers in small cell
	  lung cancer.
	\newblock \emph{Biochimica et Biophysica Acta (BBA)-Reviews on Cancer}, page
	  188552, 2021.
	
	\bibitem[Phipson et~al.(2016)Phipson, Lee, Majewski, Alexander, and
	  Smyth]{phipson2016robust}
	B.~Phipson, S.~Lee, I.~J. Majewski, W.~S. Alexander, and G.~K. Smyth.
	\newblock {Robust hyperparameter estimation protects against hypervariable
	  genes and improves power to detect differential expression}.
	\newblock \emph{Ann. Appl. Stat.}, 10\penalty0 (2):\penalty0 946, 2016.
	
	\bibitem[Quinto(2020)]{Q2020}
	B.~Quinto.
	\newblock Next-generation machine learning with spark: Covers xgboost lightgbm
	  spark nlp distributed deep learning with keras and more.
	\newblock \emph{Next-Generation Mach. Learn. with Spark Cover. XGBoost, Light.
	  Spark NLP, Distrib. Deep Learn. with Keras, More}, pages 1--355, 2020.
	
	\bibitem[Ross(2014)]{R2014}
	B.~C. Ross.
	\newblock Mutual information between discrete and continuous data sets.
	\newblock \emph{PloS one}, 9\penalty0 (2):\penalty0 e87357, 2014.
	
	\bibitem[Rudin et~al.(2021)Rudin, Brambilla, Faivre-Finn, and Sage]{RBFS:2021}
	C.~M. Rudin, E.~Brambilla, C.~Faivre-Finn, and J.~Sage.
	\newblock Small-cell lung cancer.
	\newblock \emph{Nature Reviews Disease Primers}, 7\penalty0 (1):\penalty0
	  1--20, 2021.
	
	\bibitem[Safran et~al.(2010)Safran, Dalah, Alexander, Rosen, Iny~Stein,
	  Shmoish, Nativ, Bahir, Doniger, Krug, et~al.]{SDAR2010}
	M.~Safran, I.~Dalah, J.~Alexander, N.~Rosen, T.~Iny~Stein, M.~Shmoish,
	  N.~Nativ, I.~Bahir, T.~Doniger, H.~Krug, et~al.
	\newblock Genecards version 3: the human gene integrator.
	\newblock \emph{Database}, 2010, 2010.
	
	\bibitem[Salmena et~al.(2011)Salmena, Poliseno, Tay, Kats, and
	  Pandolfi]{SPTKP:2011}
	L.~Salmena, L.~Poliseno, Y.~Tay, L.~Kats, and P.~P. Pandolfi.
	\newblock {A ceRNA hypothesis: the Rosetta Stone of a hidden RNA language?}
	\newblock \emph{Cell}, 146\penalty0 (3):\penalty0 353--358, 2011.
	
	\bibitem[Sayyed et~al.(2021)Sayyed, Gondaliya, Bhat, Mali, Arya, Khairnar, and
	  Kalia]{SGBMAKK:2021}
	A.~A. Sayyed, P.~Gondaliya, P.~Bhat, M.~Mali, N.~Arya, A.~Khairnar, and
	  K.~Kalia.
	\newblock Role of mirnas in cancer diagnostics and therapy: A recent update.
	\newblock \emph{Current pharmaceutical design}, 2021.
	
	\bibitem[Segal et~al.(2018)Segal, Miller, and Jemal]{SMJ:2018}
	R.~Segal, K.~Miller, and A.~Jemal.
	\newblock Cancer statistics, 2018.
	\newblock \emph{CA Cancer J. Clin.}, 68:\penalty0 7--30, 2018.
	
	\bibitem[Seo et~al.(2012)Seo, Ju, Lee, Shin, Lee, Bleazard, Lee, Jung, Kim,
	  Shin, et~al.]{Seoetal:2012}
	J.-S. Seo, Y.~S. Ju, W.-C. Lee, J.-Y. Shin, J.~K. Lee, T.~Bleazard, J.~Lee,
	  Y.~J. Jung, J.-O. Kim, J.-Y. Shin, et~al.
	\newblock The transcriptional landscape and mutational profile of lung
	  adenocarcinoma.
	\newblock \emph{Geno. Res.}, 22\penalty0 (11):\penalty0 2109--2119, 2012.
	
	\bibitem[Smyth(2004)]{Smyth:2004}
	G.~Smyth.
	\newblock Linear models and empirical bayes methods for assessing differential
	  expression in microarray experiments.
	\newblock \emph{Stat. Appl. Genet. Mol. Biol.}, 3:\penalty0 3, 2004.
	
	\bibitem[Sung et~al.(2021)Sung, Ferlay, Siegel, Laversanne, Soerjomataram,
	  Jemal, and Bray]{SFSLSJB:2021}
	H.~Sung, J.~Ferlay, R.~L. Siegel, M.~Laversanne, I.~Soerjomataram, A.~Jemal,
	  and F.~Bray.
	\newblock Global cancer statistics 2020: Globocan estimates of incidence and
	  mortality worldwide for 36 cancers in 185 countries.
	\newblock \emph{CA: a cancer journal for clinicians}, 71\penalty0 (3):\penalty0
	  209--249, 2021.
	
	\bibitem[Tay et~al.(2014{\natexlab{a}})Tay, Karreth, and Pandolfi]{TKP:2014}
	Y.~Tay, F.~A. Karreth, and P.~P. Pandolfi.
	\newblock {Aberrant ceRNA activity drives lung cancer}.
	\newblock \emph{Cell Res.}, 24\penalty0 (3):\penalty0 259, 2014{\natexlab{a}}.
	
	\bibitem[Tay et~al.(2014{\natexlab{b}})Tay, Rinn, and Pandolfi]{TRP:2014}
	Y.~Tay, J.~Rinn, and P.~P. Pandolfi.
	\newblock {The multilayered complexity of ceRNA crosstalk and competition}.
	\newblock \emph{Nature}, 505\penalty0 (7483):\penalty0 344, 2014{\natexlab{b}}.
	
	\bibitem[Vergoulis et~al.(2011)Vergoulis, Vlachos, Alexiou, Georgakilas,
	  Maragkakis, Reczko, Gerangelos, Koziris, Dalamagas, and
	  Hatzigeorgiou]{Vergoulisetal:2011}
	T.~Vergoulis, I.~S. Vlachos, P.~Alexiou, G.~Georgakilas, M.~Maragkakis,
	  M.~Reczko, S.~Gerangelos, N.~Koziris, T.~Dalamagas, and A.~G. Hatzigeorgiou.
	\newblock {TarBase 6.0: capturing the exponential growth of miRNA targets with
	  experimental support}.
	\newblock \emph{Nucl. Acids Res.}, 40\penalty0 (D1):\penalty0 D222--D229, 2011.
	
	\bibitem[Wei et~al.(2018)Wei, Guo, Liao, Chen, Wang, Ni, and Liang]{WGLCW2018}
	H.-T. Wei, E.-N. Guo, X.-W. Liao, L.-S. Chen, J.-L. Wang, M.~Ni, and C.~Liang.
	\newblock Genome-scale analysis to identify potential prognostic microrna
	  biomarkers for predicting overall survival in patients with colon
	  adenocarcinoma.
	\newblock \emph{Oncology reports}, 40\penalty0 (4):\penalty0 1947--1958, 2018.
	
	\bibitem[Wu et~al.(2019)Wu, Tsai, Lien, Kuo, and Hung]{WTLKH:2019}
	K.~L. Wu, Y.~M. Tsai, C.~T. Lien, P.~L. Kuo, and J.~Y. Hung.
	\newblock The roles of microrna in lung cancer.
	\newblock \emph{International Journal of Molecular Sciences}, 20\penalty0 (7),
	  2019.
	
	\bibitem[Xiao et~al.(2008)Xiao, Zuo, Cai, Kang, Gao, and Li]{XZCKGL:2008}
	F.~Xiao, Z.~Zuo, G.~Cai, S.~Kang, X.~Gao, and T.~Li.
	\newblock {miRecords: an integrated resource for microRNA--target
	  interactions}.
	\newblock \emph{Nucl. Acids Res.}, 37\penalty0 (suppl\_1):\penalty0 D105--D110,
	  2008.
	
	\bibitem[Xie et~al.(2013)Xie, Ding, Han, and Wu]{XDHW2013}
	B.~Xie, Q.~Ding, H.~Han, and D.~Wu.
	\newblock mircancer: a microrna--cancer association database constructed by
	  text mining on literature.
	\newblock \emph{Bioinformatics}, 29\penalty0 (5):\penalty0 638--644, 2013.
	
	\bibitem[Xin et~al.(2020)Xin, Cao, Zhao, Lv, Qiu, Li, Wang, Fang, and
	  Jia]{XCZLQ2020}
	G.~Xin, X.~Cao, W.~Zhao, P.~Lv, G.~Qiu, Y.~Li, B.~Wang, B.~Fang, and Y.~Jia.
	\newblock Microrna expression profile and tnm staging system predict survival
	  in patients with lung adenocarcinoma.
	\newblock \emph{Mathematical Biosciences and Engineering: MBE}, 17\penalty0
	  (6):\penalty0 8074--8083, 2020.
	
	\bibitem[Yan et~al.(2018)Yan, Cai, Guan, He, Zhang, Guo, Huang, Li, Li, Gu,
	  et~al.]{YCGHZ2018}
	H.~Yan, H.~Cai, Q.~Guan, J.~He, J.~Zhang, Y.~Guo, H.~Huang, X.~Li, Y.~Li,
	  Y.~Gu, et~al.
	\newblock Individualized analysis of differentially expressed mirnas with
	  application to the identification of mirnas deregulated commonly in lung
	  cancer tissues.
	\newblock \emph{Briefings in bioinformatics}, 19\penalty0 (5):\penalty0
	  793--802, 2018.
	
	\bibitem[Yang et~al.(2020)Yang, Yin, Shi, and Qian]{YYSQ:2020}
	Z.~Yang, H.~Yin, L.~Shi, and X.~Qian.
	\newblock A novel microrna signature for pathological grading inlung
	  adenocarcinoma based on tcga and geo data.
	\newblock \emph{International Journal of Molecular Medicine}, 45\penalty0 (5),
	  2020.
	
	\bibitem[Yu et~al.(2019)Yu, Yong, Kim, Choi, Jung, Kim, Seo, Lee, Baek, Lee,
	  et~al.]{YYKCJ2019}
	N.~Yu, S.~Yong, H.~K. Kim, Y.-L. Choi, Y.~Jung, D.~Kim, J.~Seo, Y.~E. Lee,
	  D.~Baek, J.~Lee, et~al.
	\newblock Identification of tumor suppressor mirnas by integrative mirna and
	  mrna sequencing of matched tumor--normal samples in lung adenocarcinoma.
	\newblock \emph{Molecular oncology}, 13\penalty0 (6):\penalty0 1356--1368,
	  2019.
	
	\bibitem[Zhang et~al.(2021)Zhang, Ma, Zhai, Chen, Li, Shang, Zhang, Gao, Yang,
	  Li, et~al.]{ZMZCL2021}
	X.~Zhang, L.~Ma, L.~Zhai, D.~Chen, Y.~Li, Z.~Shang, Z.~Zhang, Y.~Gao, W.~Yang,
	  Y.~Li, et~al.
	\newblock Construction and validation of a three-microrna signature as
	  prognostic biomarker in patients with hepatocellular carcinoma.
	\newblock \emph{International journal of medical sciences}, 18\penalty0
	  (4):\penalty0 984, 2021.
	
	\bibitem[Zhao et~al.(2021)Zhao, Feng, and Tang]{ZFT2021}
	M.~Zhao, J.~Feng, and L.~Tang.
	\newblock Competing endogenous rnas in lung cancer.
	\newblock \emph{Cancer Biology \& Medicine}, 18\penalty0 (1):\penalty0 1, 2021.
	
	\bibitem[Zhao et~al.(2018)Zhao, Wang, Wu, Yan, Wu, Wang, Yang, and
	  Shao]{ZWWYWWYS:2018}
	Y.~Zhao, H.~Wang, C.~Wu, M.~Yan, H.~Wu, J.~Wang, X.~Yang, and Q.~Shao.
	\newblock {Construction and investigation of lncRNA-associated ceRNA regulatory
	  network in papillary thyroid cancer}.
	\newblock \emph{Oncol. Rep.}, 39\penalty0 (3):\penalty0 1197--1206, 2018.
	
	\bibitem[Zhong et~al.(2020)Zhong, Golpon, Zardo, and Borlak]{ZGZB:2020}
	S.~Zhong, H.~Golpon, P.~Zardo, and J.~Borlak.
	\newblock mirnas in lung cancer. a systematic review identifies predictive and
	  prognostic mirna candidates for precision medicine in lung cancer.
	\newblock \emph{Translational Research}, 2020.
	
	\bibitem[Zhong et~al.(2021)Zhong, Golpon, Zardo, et~al.]{ZGZO2021}
	S.~Zhong, H.~Golpon, P.~Zardo, et~al.
	\newblock mirnas in lung cancer. a systematic review identifies predictive and
	  prognostic mirna candidates for precision medicine in lung cancer.
	\newblock \emph{Translational Research}, 230:\penalty0 164--196, 2021.
	
	\end{thebibliography}

\end{document}